\documentclass[12pt]{article}
\usepackage{epsfig}\parskip 5pt plus 1pt
\newcommand{\raw}{\rightarrow}

\newcommand{\be}{\begin{equation}}
\newcommand{\ee}{\end{equation}}
\newcommand{\bea}{\begin{eqnarray}}
\newcommand{\eea}{\end{eqnarray}}

\newcommand{\He}{\ensuremath{^6{\mathrm{He}\,}}}
\newcommand{\Ne}{\ensuremath{^{18}{\mathrm{Ne}\,}}}

\def\nue{\ensuremath{\nu_{e}}}
\def\nubare{\ensuremath{\overline{\nu}_{e}}}

\def\simge{\mathrel{%
   \rlap{\raise 0.511ex \hbox{$>$}}{\lower 0.511ex \hbox{$\sim$}}}}
\def\simle{\mathrel{
   \rlap{\raise 0.511ex \hbox{$<$}}{\lower 0.511ex \hbox{$\sim$}}}}

\begin{document}
\thispagestyle{empty}
\begin{flushright}
{IFT-UAM/CSIC-04-30}
\end{flushright}
\vspace*{1cm}
\begin{center}
{\Large{\bf Study of the eightfold degeneracy with a standard
$\beta$-Beam and a Super-Beam facility} }\\
\vspace{.5cm}
A. Donini$^{\rm a}$, E. Fernandez-Martinez$^{\rm a}$, P. Migliozzi$^{\rm b}$,
S. Rigolin$^{\rm a}$ and L. Scotto Lavina$^{\rm c}$ \\
\vspace*{1cm}
$^{\rm a}$ I.F.T. and Dep. F\'{\i}sica Te\'{o}rica, U.A.M., E-28049, Madrid, Spain \\
$^{\rm b}$ I.N.F.N., Sezione di Napoli, I-80126, Napoli, Italy \\
$^{\rm c}$ Dip. Fisica, Universit\`{a} di Napoli``Federico II'' and I.N.F.N., I-80126,
Napoli, Italy
\end{center}

\vspace{.3cm}
\begin{abstract}
\noindent
The study of the eightfold degeneracy at a neutrino complex that
includes a standard $\beta$-Beam and a Super-Beam facility is
presented for the first time in this paper. The scenario where the
neutrinos are sent toward a Megaton water Cerenkov detector located
at the Fr\'{e}jus laboratory (baseline 130 Km) is exploited.  The
performance in terms of sensitivity for measuring the continuous
($\theta_{13}$ and $\delta$) and discrete ($\mbox{sign} [ \Delta
m^2_{23} ]$ and $\mbox{sign} [\tan (2\theta_{23}) ]$) oscillation
parameters for the $\beta$-Beam and Super-Beam alone, and for their
combination has been studied. A brief review of the present
uncertainties on the neutrino and antineutrino cross-sections is also
reported and their impact on the discovery potential discussed.
\end{abstract}

\vspace*{\stretch{2}}
\begin{flushleft}
\end{flushleft}

\newpage

\section{Introduction}
\label{introduction}
In the past years the hypothesis of neutrino oscillations has been
strongly confirmed in the atmospheric~\cite{atmo},
accelerator~\cite{Ahn:2002up}, solar~\cite{solar} and
reactor~\cite{reactor} sectors. If we do not consider the claimed
evidence from the LSND experiment~\cite{lsnd}, that must be confirmed
or excluded by the ongoing MiniBooNE experiment~\cite{boone},
oscillations in the leptonic sector can be accommodated in the three
family Pontecorvo-Maki-Nakagawa-Sakata (PMNS) mixing
matrix~\cite{neutrino_osc}. Therefore, the next steps on the way of a
full understanding of neutrino oscillations are: 


\begin{itemize}
  \item confirm the source of atmospheric neutrino oscillations, i. e. observe directly
        the $\nu_\mu\rightarrow\nu_\tau$ oscillation;
  \item perform precision measurements of the angles $\theta_{12}$ and
        $\theta_{23}$ and of the mass differences $\mid \Delta
        m^2_{12}\mid$ and $\mid\Delta m^2_{23}\mid$;
  \item measure the sign of the atmospheric mass difference, $\Delta m^2_{23}$;
  \item measure the remaining parameters of the PMNS mixing matrix:
        $\theta_{13}$ (for which only an upper limit exists so
        far~\cite{chooz}) and the leptonic CP violating phase $\delta$
        (that is still completely unknown).
\end{itemize}

With the aim to perform the above measurements, over recent years
there has been a marked growth of interest in the development of new
neutrino sources: conventional neutrino beams from pion and kaon
decay, but with a more intense flux (Super-Beams); neutrino beams from
muon decays (Neutrino Factories); neutrino beams from the decay of
intense beams of $\beta$-unstable heavy ions ($\beta$-Beams). For a
comprehensive review of future neutrino sources we refer
to~\cite{Apollonio:2002en} and references therein.

In this paper we focus on a CERN-based neutrino complex including a
$\beta$-Beam, that could leverage existing facilities at CERN and
complement the EURISOL physics program~\cite{betabeams_moriond}, and a
Super-Beam based on an intense proton driver (the SPL). Although the
possibility to exploit higher $\gamma$ $\beta$-Beams has been put
forward (see for
example~\cite{Burguet-Castell:2003vv,Terranova:2004hu}), we consider
here only the configuration of the $\beta$-Beam where $\nu_e$
($\bar{\nu}_e$) are produced by $^{18}$Ne ($^{6}$He)
ions~\cite{Bouchez:2003fy} that are accelerated by the SPS up to
$\gamma\sim100$ ($\gamma\sim60$), respectively (standard
$\beta$-Beam).  These $\gamma$ values have been chosen in order to
tune the neutrino/antineutrino mean energy in such a way that the
maximum of the atmospheric neutrino oscillation lies at a distance of
around 100 Km, i.e the distance from CERN to Fr\'{e}jus.

A first estimate of the potentiality of a CERN to Fr\'{e}jus based neutrino complex
was given in Ref.~\cite{Bouchez:2003fy}. However, in that study only the intrinsic
degeneracy\footnote{i.e. the sign of the atmospheric mass difference $\Delta m^2_{23}$
is assumed to be positive and $\theta_{23}=45^\circ$.} was taken into account and only
the very peculiar value $\delta=90^\circ$ was considered.
In Ref.~\cite{Burguet-Castell:2003vv} a more thoughtful analysis, on
both experimental and theoretical issues, appears even if again only
positive $\Delta m^2_{23}$ and $\theta_{23}=45^\circ$ were considered,
the main interest of the paper being the study of higher $\gamma$
setups.

The Super-Beam envisaged at the CERN neutrino complex studied in this
paper is based on the planned SPL of 4 MW power described in
Ref.~\cite{Gomez-Cadenas:2001eu}. Similar projects have been also
proposed in Japan and USA, and carefully studied by several
authors~\cite{allSB}, with neutrinos energies around 1-2 GeV. The
Super-Beam studied in this paper has an average neutrino energy around
0.25 GeV to match the CERN-Fr\'{e}jus distance. A comprehensive
analysis of the CERN-based Super-Beam potential, although based on old
fluxes and considering only the intrinsic degeneracy, can be found
in~\cite{Gomez-Cadenas:2001eu}.

In this paper we study for the first time the complete eightfold degeneracy for the
CERN neutrino complex, i.e. the $\beta$-Beam and the Super-Beam either separately or
together. In Section~\ref{theo}, the neutrino oscillation formalism is introduced and
the eightfold degeneracy discussed from a theoretical point of view. The main features
of the neutrino complex at CERN are discussed in Section~\ref{beams}, while the set of
cross-sections used in this paper are briefly summarized in Section~\ref{cross}. In
particular, we compared these cross-sections with other calculations and briefly
comment the present status of the cross-section knowledge for neutrino energies below
1 GeV. Finally we give our results on the sensitivity calculation both for the
$\theta_{13}$ angle and the CP violating phase $\delta$.

\section{The eightfold degeneracy}
\label{theo}

In Ref.~\cite{Burguet-Castell:2001ez} it has been noticed that the
appearance probability $P_{\alpha \beta}$ obtained for neutrinos at a
fixed energy and baseline with input parameter
($\bar\theta_{13},\bar\delta$) has no unique solution. Indeed, the
equation \be
\label{eq:equi0}
P_{\alpha\beta} (\bar\theta_{13},\bar\delta) = P_{\alpha\beta}
(\theta_{13},\delta)
\ee
has a continuous number of solutions. The locus of the ($\theta_{13},\delta$) plane 
satisfying this equation is called ``equiprobability curve''. Considering the 
equiprobability curves for neutrinos and antineutrinos with the same energy (and the 
same input parameters), the following system of equations ($\pm$ referring to 
neutrinos and antineutrinos respectively)
\be
\label{eq:equi1}
P^\pm_{\alpha \beta} (\bar \theta_{13},\bar \delta) = P^\pm_{\alpha \beta}
(\theta_{13}, \delta)
\ee
has two intersections: the input pair ($\bar \theta_{13},\bar \delta$)
and a second, energy dependent, point. This second intersection
introduces an ambiguity in the measurement of the physical values of
$\theta_{13}$ and $\delta$: the so-called {\it intrinsic clone}
solution. Knowing the two probabilities of Eq.~(\ref{eq:equi1}) is
consequently not enough for solving the intrinsic degeneracy. One
needs to add more information.

Unfortunately the appearance of the intrinsic degeneracy is only a
part of the ``clone problem''. As it was made clear in
\cite{Minakata:2001qm,Fogli:1996pv,Barger:2001yr}, two other sources
of ambiguities arise due to the present (and near future) ignorance of
the sign of the atmospheric mass difference,
$s_{atm}=~\mbox{sign}[\Delta m^2_{23}]$ and the $\theta_{23}$ octant,
namely $s_{oct}=~\mbox{sign} [\tan(2\theta_{23})]$. These two discrete
variables assume the values $\pm 1$, depending on the physical
assignments of the $\Delta m^2_{23}$ sign ($s_{atm}=1$ for
$m_3^2>m_2^2$ and $s_{atm}=-1$ for $m_3^2<m_2^2$) and of the
$\theta_{23}$-octant ($s_{oct}=1$ for $\theta_{23}<\pi/4$ and
$s_{oct}=-1$ for $\theta_{23}>\pi/4$). As a consequence, future
experiments will have as ultimate goal the measure of the two
continuous variables $\theta_{13}$ and $\delta$ plus the two discrete
variables $s_{atm}$ and $s_{oct}$.

Moreover it should be noticed that experimental results are not given
in terms of oscillation probabilities but of number of charged leptons
observed in a specific detector. It has been
noticed~\cite{Donini:2003vz} that clones location calculated starting
from the probability or the number of events can be significantly
different. We must therefore integrate the oscillation probability
over the neutrino flux, the $\nu N$ cross-section and the detector
efficiency $\epsilon (E_\mu)$.  From these considerations it follows
that Eq.~(\ref{eq:equi1}) should be more correctly replaced by: \be
\label{eq:ene0int}
N^\pm_{\beta} (\bar \theta_{13},\bar \delta; \bar s_{atm},\bar s_{oct}) =
N^\pm_{\beta} (\theta_{13},\delta; s_{atm}=\bar s_{atm}; s_{oct}=\bar
s_{oct})\, .
\ee
In Eq.~(\ref{eq:ene0int}) we have implicitly assumed to know the right
sign and the right octant for the atmospheric mass difference and
angle. As these quantities are unknown (and presumably they will still
be unknown at the time of the neutrino facilities considered in this
paper) the following systems of equations should be considered as
well:
\bea
\label{eq:ene0sign}
N^\pm_{\beta}(\bar \theta_{13}, \bar \delta; \bar s_{atm}, \bar s_{oct} )&=&
N^\pm_{\beta} ( \theta_{13}, \delta; s_{atm} = -\bar s_{atm}, s_{oct} = \bar
s_{oct})\\
\label{eq:ene0t23}
N^\pm_{\beta}(\bar \theta_{13}, \bar \delta; \bar s_{atm}, \bar s_{oct}) &=&
N^\pm_{\beta} ( \theta_{13},  \delta; s_{atm} = \bar s_{atm}, s_{oct} = -\bar
s_{oct})\\
\label{eq:ene0t23sign}
N^\pm_{\beta}(\bar \theta_{13}, \bar \delta; \bar s_{atm}, \bar s_{oct} )&=&
N^\pm_{\beta} ( \theta_{13},  \delta; s_{atm} = -\bar s_{atm}, s_{oct} =
-\bar s_{oct})
\eea
Solving the four systems of
Eqs.~(\ref{eq:ene0int})-(\ref{eq:ene0t23sign}) will result in
obtaining the true solution plus additional {\it clones} to form an
eightfold degeneracy~\cite{Barger:2001yr}. These eight solutions are
respectively:
\begin{itemize}
\item the true solution and its {\em intrinsic clone}, obtained solving
the system
      in Eq.~(\ref{eq:ene0int});
\item the $\Delta m^2_{23}$-sign clones (hereafter called {\em sign
clones}) of the
      true and intrinsic solution, obtained solving the system in
Eq.~(\ref{eq:ene0sign});
\item the $\theta_{23}$-octant clones (hereafter called {\em octant
clones}) of the
      true and intrinsic solution, obtained solving the system in
Eq.~(\ref{eq:ene0t23});
\item the $\Delta m^2_{atm}$-sign $\theta_{23}$-octant clones (hereafter
called
      {\em mixed clones}) of the true and intrinsic solution, obtained
solving the system
      in Eq.~(\ref{eq:ene0t23sign}).
\end{itemize}

A complete description of the clone location has been done in
~\cite{Donini:2003vz} and we refer to that article for all the
theoretical details. In this paper we are interested in presenting a
detailed analysis of a concrete experimental facility, that we are
going to describe in the following Sections.


\section{Neutrino beam facilities at CERN}
\label{beams}

In this Section we summarize some of the technical details of the two considered 
neutrino beams, the standard $\beta$-Beam (Section~\ref{bbeam}) and the Super-Beam 
(Section~\ref{sbeam}). Both beams are directed from CERN toward the underground 
Fr\'{e}jus laboratory, where it has been proposed to locate a 1 Megaton UNO-like 
\cite{Jung:1999jq} water Cerenkov detector with a 440 Kt fiducial mass. The considered 
baseline is thus L = 130 Km for both beams. Therefore, in order to be at the maximum of 
the atmospheric neutrino oscillations, the peaks of the energy spectra have been chosen 
of order few hundred MeV.

\subsection{The $\beta$-Beam}
\label{bbeam}

The $\beta$-Beam concept was first introduced in
Ref.~\cite{Zucchelli:sa}.  It involves producing a beam of
$\beta$-unstable heavy ions, accelerating them to some reference
energy, and allowing them to decay in the straight section of a
storage ring, resulting in a very intense neutrino beam.  Two ions
have been identified as ideal candidates: $^6$He, to produce a pure
$\bar{\nu}_e$ beam, and $^{18}$Ne, to produce a $\nu_e$ beam. The
``golden'' \cite{Cervera:2000kp} sub-leading transitions
$\nu_e\rightarrow \nu_\mu$ and ${\bar\nu}_e \rightarrow {\bar\nu}_\mu$
can be measured through the appearance of muons in a distant detector.

%
\begin{figure}[t!]
\vspace{-0.5cm}
\begin{center}
\begin{tabular}{c}
\hspace{-0.3cm} \epsfxsize10cm\epsffile{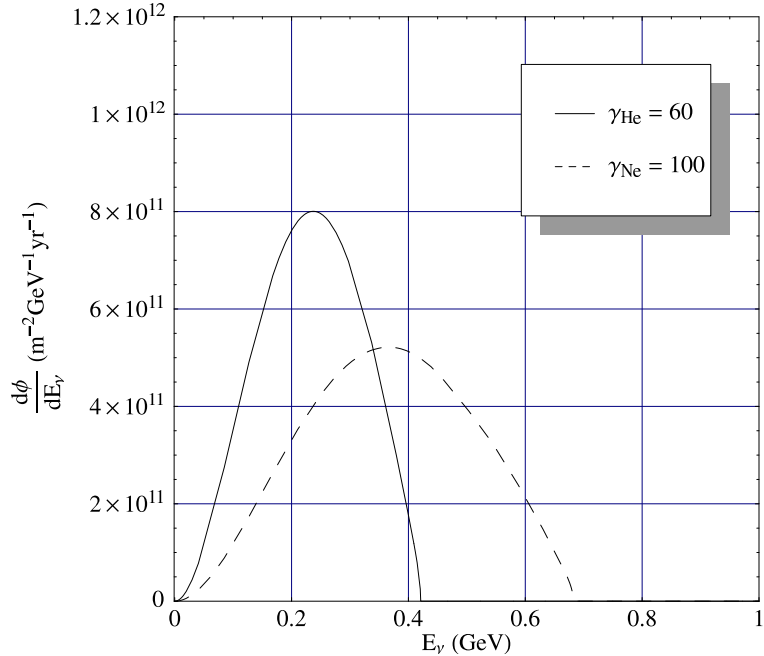}\\
\end{tabular}
\caption{\it \label{fig:fluxes}
$\beta$-Beam fluxes at the Fr{\'{e}}jus location (130 km baseline) as a function of
the neutrino energy for the two specific $\gamma$ values shown in the legend.}
\end{center}
\end{figure}

The neutrino beam energy depends on the $\gamma$ of the parent ions in
the decay ring. For the scenario considered in this paper the $\gamma$
ratio for the two ions has been fixed to be $\gamma(^6{\rm
He})/\gamma(^{18}{\rm Ne})=3/5$~\cite{Bouchez:2003fy}. This constraint
comes from the request to accelerate in the same accelerator at the
same time ions with different atomic mass. The optimal $\gamma$ values
that match the CERN-Fr\'{e}jus distance have been found to be
$\gamma(\He)=60$ and $\gamma(\Ne)=100$. The mean neutrino energies of
the \nubare, \nue\ beams corresponding to this configuration are
0.23~GeV and 0.37~GeV, respectively. On the other hand the energy
resolution is very poor at these energies, given the influence of
Fermi motion and other nuclear effects. Therefore, in the following
all the sensitivities are computed for a counting experiment with no
energy cuts.

A flux of $2.9 {\times} 10^{18}$ \He\ decays/year and
$1.1{\times}10^{18}$ \Ne\ decays/year, as discussed in
Ref.~\cite{Bouchez:2003fy}, will be assumed. Fig.~\ref{fig:fluxes}
shows the $\beta$-Beam neutrino fluxes computed at the 130 Km
baseline, keeping $m_e \neq 0$, following the formulas derived in
Ref.~\cite{Burguet-Castell:2003vv}, while in
Ref.~\cite{Bouchez:2003fy} the fluxes are calculated in the $m_e=0$
approximation. Be aware of the fact that even if $m_e$ effects seem
negligible, their inclusion could be sizable due to the dramatic
cross-section suppression of low energy neutrinos\footnote{For example
we checked that by using the cross-sections discussed in
Section~\ref{cross}, the rate, for both neutrinos and antineutrinos,
computed in the $m_e=0$ approximation is about 20\% smaller than in
the case $m_e \neq 0$.}. Furthermore, in our calculations we take into
account the fact that the $^{18}$Ne has three different decay modes,
each with a different end-point energy, see Table~\ref{betabeam}.

\begin{table}
\begin{center}
\begin{tabular}{|c|c|c|} \hline \hline
   Element & End-Point (MeV) & Decay Fraction \\ \hline
            & 34.114 & 92.1\% \\
  $^{18}$Ne & 23.699 & 7.7\% \\
            & 17.106 & 0.2\% \\ \hline
 $^{6}$He   & 35.078 & 100\% \\ \hline
\hline
\end{tabular}
\caption{\it \label{betabeam} $^{18}$Ne and $^{6}$He $\beta$-decay channels and relative
end-point energies from~\cite{betadecays}.}
\end{center}
\end{table}

\subsection{The Super-Beam}
\label{sbeam}

The Super-Beam is a conventional neutrino beam, but with a higher
proton intensity. Therefore, it has the advantages of a high intensity
flux and of a well proved technology. On the other hand its
composition ($\nu_\mu$ main component, if $\pi^+$ are focused, plus a
small admixture of $\bar{\nu}_\mu$, $\nu_e$ and $\bar{\nu}_e$) is
affected by large systematic uncertainties that limit the sensitivity
in searching for neutrino oscillations mainly
$\nu_\mu\rightarrow\nu_e$.

\begin{figure}[t!]
\vspace{-0.5cm}
\begin{center}
\begin{tabular}{c}
\hspace{-0.3cm} \epsfxsize11cm\epsffile{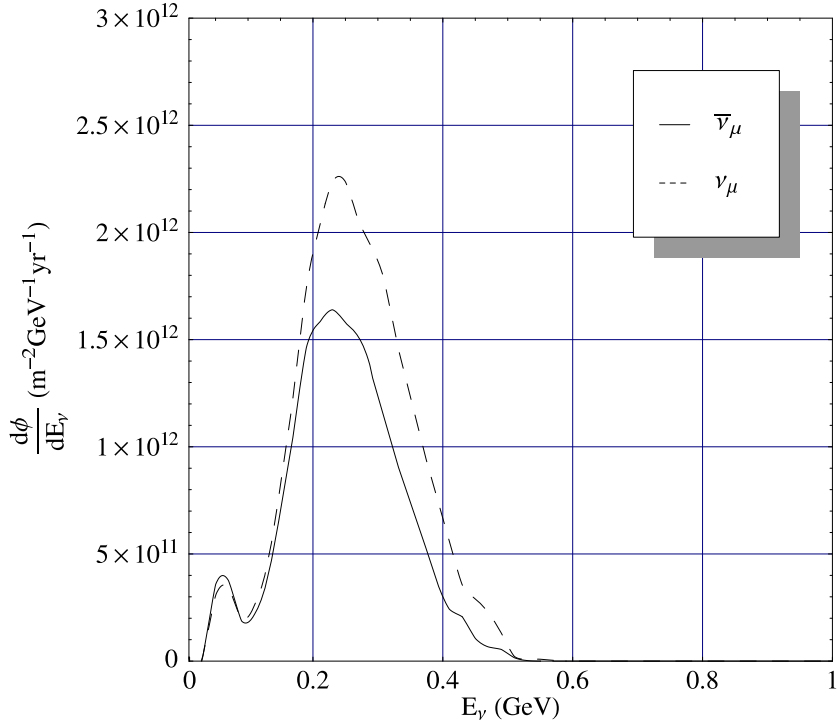}\\
\end{tabular}
 \caption{\it
SPL Super-Beam fluxes at the Fr{\'{e}}jus location (130 km baseline) as a
function of the neutrino energy~\cite{gilardoni}. }
\label{fig:sbfluxes}
\end{center}
\end{figure}

As baseline for this work we consider a Super-Beam based on a 2.2 GeV
proton beam of 4~MW power SPL, described in
Ref.~\cite{Gomez-Cadenas:2001eu}. The predicted energy spectra and
fluxes for the main components have been computed starting from a full
simulation of the neutrino beamline~\cite{gilardoni}, assuming a decay
tunnel length of 60~m. The neutrino fluxes expected at the Fr\'{e}jus
location are shown in Fig.~\ref{fig:sbfluxes}. The average energy of
the neutrino and antineutrino beams is 0.27 GeV and 0.25 GeV,
respectively.

%
%
%


%
\begin{figure}[t!]
\vspace{-0.5cm}
\begin{center}
\begin{tabular}{c}
\hspace{-0.3cm} \epsfxsize10cm\epsffile{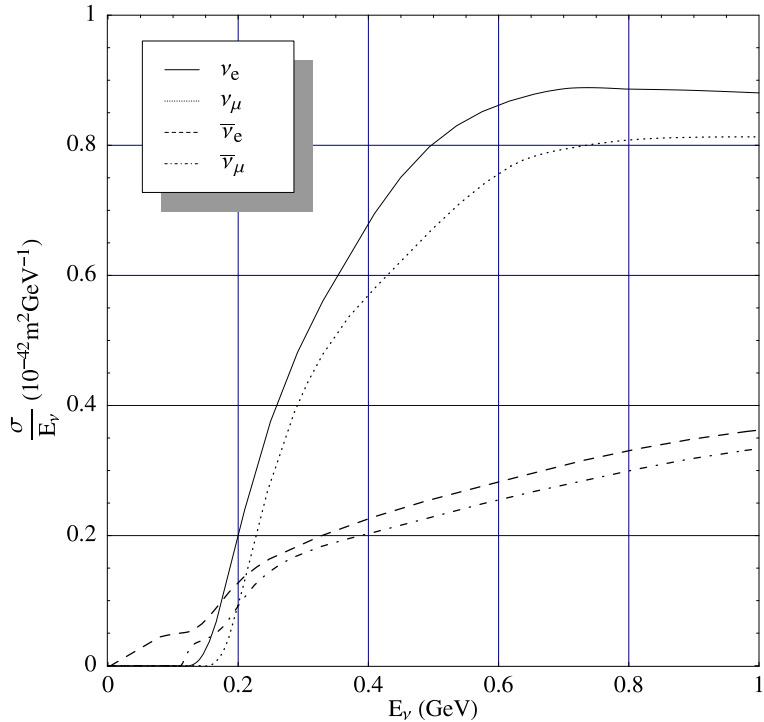}\\
\end{tabular}
\caption{\it Cross-sections on water as a function of the neutrino energy~\cite{lipari}.  }
\label{fig:crossec}
\end{center}
\end{figure}

\section{Neutrino cross-sections}
\label{cross}

The present knowledge of the neutrino and antineutrino cross-sections
for energies below 1~GeV is very poor~\cite{Zeller:2003ey}: either there are 
very few data, as for neutrinos, or there are no data available at all, as for 
antineutrinos. On top of that, the few available data have not been taken on 
water, the target we are interested in, and the extrapolation from different 
nuclei is complicated by nuclear effects that at the considered energies play 
an important role. 
For this calculation we adopted the cross-sections on water shown in
Fig.~\ref{fig:crossec}~\cite{lipari}. Notice the difference between
the $\nu_e N$ and $\bar \nu_e N$ cross-sections: the former, being an
interaction between the $\nu_e$ and a neutron inside the oxygen
nucleus, is affected by nuclear effects and thus shows a threshold
energy. The latter is mainly a $\bar \nu_e$ interaction with the
protons of the two hydrogens, approximately free. This effect,
although less pronounced, is visible also for $\nu_\mu$ and for
$\bar\nu_\mu$. This feature is quite relevant for
neutrino/antineutrino of hundreds of MeV energy, region where
different cross-sections can easily differ by a factor 2. Compare for
example our Fig.~\ref{fig:crossec} with Fig.~3 of
Ref.~\cite{Burguet-Castell:2003vv} where NUANCE
cross-sections~\cite{Casper:2002sd} are plotted. These differences can
explain the (sometimes relevant) discrepancies within the numbers of
charged-current interactions and oscillated events predicted in
different analyzes. Be aware that there are other nuclear effects
(see~\cite{Serreau:2004kx} and references therein) not included yet in
any of the available calculations that could play an important effect
at the cross-section threshold energy.

With our calculations the expected number of charged-current
events with a $\beta$-Beam without oscillations per kton-year is 30.3
and 4.4 for $\nu_e$ and $\bar\nu_e$, respectively. While in 
Ref.~\cite{Burguet-Castell:2003vv}, they quote 32.8 and 4.7 for $\nu_e$ and 
$\bar\nu_e$, respectively. We verify, that using the NUANCE cross-sections  
we are able to reproduce exactly their results as 
well as those of Ref.~\cite{Bouchez:2003fy}. As far as the Super-Beam 
is concerned, the expected number of charged-current events with a 
Super-Beam without oscillations per kton-year is 27.6 and 7.2 for 
$\nu_\mu$ and $\bar\nu_\mu$, respectively. We verify, that using
the NUANCE cross-sections we are able to reproduce within 5\% the 
results of Ref.~\cite{Mezzetto:2003mm}, while we still found a discrepancy 
of more than 30\% with those quoted in Ref.~\cite{Bouchez:2003fy}.

Besides the absolute value of the cross-sections, another important
unknown is their shape. Indeed, as it will be discussed later, some of
the backgrounds have a neutrino energy threshold. Therefore, the
expected background strongly depends on the adopted model.

At the time the neutrino complex would become operational the
cross-sections will be measured precisely. However, nowadays we have
the problem to compute the physics potential of a facility having in
mind that the expected number of signal and background events strongly
depend on the adopted calculation.

A facility where the neutrino fluxes are known with great precision is
the ideal place where to measure neutrino cross-sections. At a
$\beta$-Beam the neutrino fluxes are completely defined by the parent
ions $\beta$-decay properties and by the number of ions in the decay
ring. A close detector of $\sim 1~$Kton placed at a distance of about
1~Km from the decay ring could then measure the relevant neutrino
cross-sections.  Furthermore the $\gamma$ factor of the accelerated
ions can be varied.  In particular a scan can be initiated below the
background production threshold, allowing a precise measurement of the
cross-sections for resonant processes.

Unfortunately, there are no studies available on the ultimate precision
achievable at future facilities in measuring the charged-current
cross-sections.  A first attempt to estimate the ultimate systematic
error achievable in the cross-section measurements at future
facilities was given in~\cite{Apollonio:2002en}. By assuming the
$\beta$-Beam complex described before a systematic error of 2\% was
estimated.


\section{$\theta_{13}$ and $\delta$ sensitivity}

In this Section we present our results for the sensitivity to
$\theta_{13}$ and $\delta$ at the CERN-based standard $\beta$-Beam and
SPL Super-Beam.  The sensitivity to the $\theta_{13}$ and $\delta$
parameters has been evaluated using the following reference values for
the solar and atmospheric parameters: \mbox{$\Delta m_{12}^2
=7.3{\times} 10^{-5} {\rm eV}^2$},
\mbox{$\theta_{12}=35^\circ$},\mbox{$\Delta m_{23}^2=2.5{\times}
10^{-3} {\rm eV}^2$} and \mbox{$\theta_{23} = 40^\circ$}.

We proceed first summarizing the backgrounds for both the facilities
(Sections~\ref{sec:bbback}-\ref{sec:sbback}) and then studying the
$\beta$-Beam and the Super-Beam separately and in combination
(Section~\ref{sec:sens}).

\subsection{Signal and background at a standard $\beta$-Beam}
\label{sec:bbback}

\begin{figure}[t!]
\vspace{-0.5cm}
\begin{center}
\begin{tabular}{c}
\hspace{-0.3cm} \epsfxsize10cm\epsffile{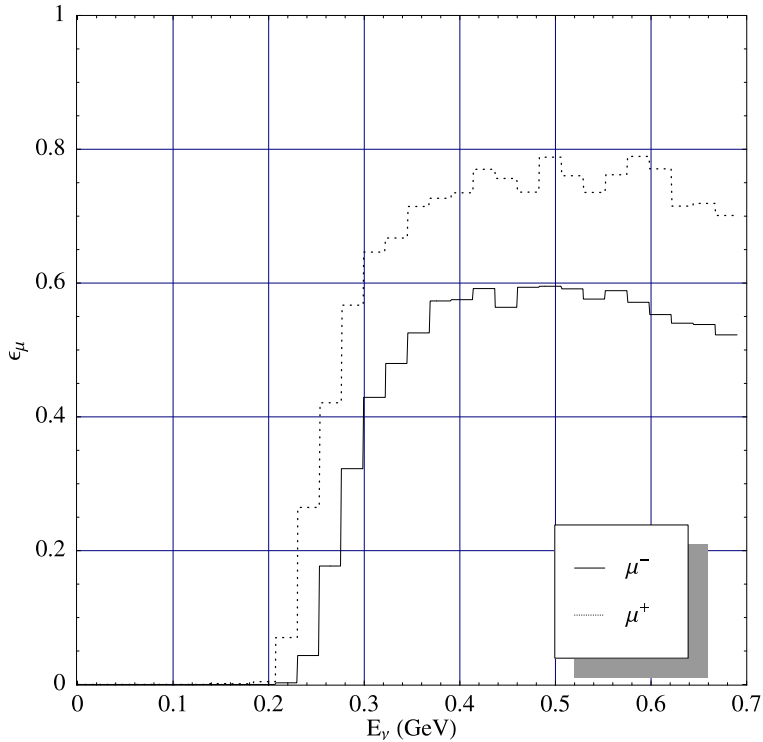}\\
\end{tabular}
\caption{\it
Reconstruction efficiency as a function of the true neutrino
energies for $^6$He ($\bar{\nu}_e$) and $^{18}$Ne (${\nu}_e$) in
water \cite{Burguet-Castell:2003vv}.
}
\label{fig:eff}
\end{center}
\end{figure}

The signal in a $\beta$-Beam looking for $\nu_e \raw \nu_\mu$
($\bar{\nu}_e \raw \bar{\nu}_\mu$) oscillations would be the
appearance of $\nu_\mu(\bar{\nu}_\mu)$ charged-current events, mainly
via quasi-elastic interactions, in a pure $\nu_e(\bar{\nu}_e)$
beam. Background rates and signal efficiencies have been studied, by
means of a full simulation based on the NUANCE
code~\cite{Casper:2002sd}, in
Refs.~\cite{Burguet-Castell:2003vv,Bouchez:2003fy}. In this paper, we
make use of those results for the beam and detector fractional
background and compute the expected number of oscillated events by
using the fluxes of Fig.~\ref{fig:fluxes}, the cross-section on water
of Fig.~\ref{fig:crossec}, the full three-families oscillation
probability in matter and the $\nu_\mu$ detection efficiency of
Fig.~\ref{fig:eff}.

%

Given its excellent purity (neither $\nu_\mu$ nor $\bar{\nu}_\mu$ are
in the initial beam) the background in a detector exploiting a
$\beta$-Beam can be generated either by inefficiencies in particle
identification, such as mis-identification of pions produced in
neutral-current single-pion resonant interactions, electrons
(positrons) mis-identified as muons, or by external sources such as
atmospheric neutrino interactions.

The pion background has a threshold at neutrino energies of about
0.45~GeV, and is highly suppressed at the $\beta$-Beam energies. The
electron background is almost completely suppressed by the request of
the detection of a delayed Michel electron following the muon track.
If a bunch length of 10~ns (which seems feasible) is assumed, this
background becomes negligible~\cite{Bouchez:2003fy}.
Moreover, out-of-spill neutrino interactions can be used to normalize
this background to the 1\% accuracy level.

The event rates for a 4400 kt-y exposure are also given in
Table~\ref{tab:beta:rates}, both for non-oscillated and oscillated
$\nu_e, \bar \nu_e$, with $\nu_e \to \nu_\mu$ oscillation probability
computed for $\theta_{13} = 10^\circ$ and $\delta = 90^\circ$, and
$\theta_{13} = 1^\circ$ and $\delta = -90^\circ,\,90^\circ$. In
Table~\ref{tab:beta:rates} the expected beam and detector backgrounds
(derived from the fractional background in
Refs.~\cite{Burguet-Castell:2003vv,Bouchez:2003fy}) are also given.

%
\begin{table}[hbtp]
\begin{center}
\begin{tabular}{|c|c|c|c|c|c|c|c|} \hline
  $\theta_{13} $ & $\delta$ &  $s_{atm}$ & $N_{\nu}$ &
  $N_{\bar{\nu}}$ & $P_{\nu_e \nu_\mu}(E=0.37)$ &
  $P_{\bar{\nu}_e \bar{\nu}_\mu}(E=0.23)$  \\ \hline \hline
      No Osc. &  &  & 133205 &  19557 &  & \\ \hline
  $10^\circ$ & $0^\circ$ & + & 2472 &  457 & 5.48$\times 10^{-2}$ & 5.17$\times 10^{-2}$\\ \hline
  $10^\circ$ & $0^\circ$ & - & 1918 &  445 & 4.35$\times 10^{-2}$ & 6.57$\times 10^{-2}$\\ \hline
       $1^\circ$ & $90^\circ$ & + & 75 &  1  & 1.85$\times 10^{-3}$ & 1.71$\times 10^{-4}$\\ \hline
       $1^\circ$ & $90^\circ$ & - & 73 &  1  & 1.79$\times 10^{-3}$ & 1.44$\times 10^{-4}$\\ \hline
       $1^\circ$ & $-90^\circ$ & + & 8 &  18 &  9.99$\times 10^{-5}$ & 3.35$\times 10^{-3}$\\ \hline
       $1^\circ$ & $-90^\circ$ & - & 8 &  20 &  9.63$\times 10^{-5}$ & 3.50$\times 10^{-3}$\\ \hline
\hline
\hline
Beam back.        & &    &  0      &  0   & & \\
\hline
Detector back.    & &    &  360     &  1  & & \\
\hline
\hline
\end{tabular}
\end{center}
\caption{\label{tab:beta:rates} \it Event rates for a 10 years
exposure at a standard $\beta$-Beam. The oscillated charged-current
events for different values of $\theta_{13},\delta$ and sign of the
atmospheric mass difference, $s_{atm}$, for both neutrinos and
antineutrinos are given. For comparison with literature we show here
the values obtained with the reference parameters but
$\theta_{23}=45^\circ$. The oscillation probabilities at the mean
neutrino/antineutrino energy (in GeV) are also shown.}
\end{table}


Finally, some comments on the overall systematic error and on the
expected background are in order.  The ultimate precision on
cross-sections achievable at future facilities is about 2\%. This
value has been also assumed in Ref.~\cite{Bouchez:2003fy} as an
overall systematic error.  However, it does not take into account
possible systematic errors on the detection efficiencies and on the
neutrino fluxes. In order to be conservative, we adopted an overall
systematic error of 5\%.  Nonetheless, we also studied the impact on
the physics potential going from 5\% to 2\%.

As far as the background is concerned, we would like to stress that it
is due to the coherent pion production process, with a threshold at
0.45 GeV. Therefore, following the arguments of Section~\ref{cross},
it strongly depends on the adopted model. We checked that our results
are stable against variations of the background in the neutrino
channel, while this is not the case for the antineutrino channel. In
absence of a full Monte Carlo simulation, in the following
calculations we adopt the central values given in
Tab.~\ref{tab:beta:rates}. The impact of an increase of this
background on the sensitivity is discussed in Section~\ref{sec:sensi}.

\subsection{Signal and background at the Super-Beam }
\label{sec:sbback}

The search for $\nu_\mu \raw \nu_e$ ($\bar{\nu}_\mu\raw\bar{\nu}_e$)
appearance with a Super-Beam is complicated by the
$\nu_e(\bar{\nu}_e)$ contamination of the beam.  Indeed, contrary to
the case of the $\beta$-Beam where the beam-induced background is
absent (see Table~\ref{tab:beta:rates}), for the Super-Beam a
significant background from $\nu_e(\bar{\nu}_e)$ charged-current
interactions must be considered, resulting in a loss of
sensitivity. In a water Cerenkov detector the appearance of a
$\nu_e(\bar{\nu}_e)$ signal is detected by exploiting the high
efficiency and purity of the detector in identifying electrons and
muons in low multiplicity interactions.

Besides the $\nu_e(\bar{\nu}_e)$ contamination of the beam, the main
sources of background are the charged-current interactions of
$\nu_\mu(\bar{\nu}_\mu)$ and the production of $\pi^0$ in
neutral-current interactions. The total background has been computed
following the fractional backgrounds given in
Ref.~\cite{Gomez-Cadenas:2001eu}.

To compute the expected number of non-oscillated and oscillated events
we have used the fluxes of Fig.~\ref{fig:sbfluxes}, the cross-section
on water of Fig.~\ref{fig:crossec}, the full three-families
oscillation probability in matter and the electron detection
efficiency computed in Ref.~\cite{Gomez-Cadenas:2001eu}: 70.7\% and
67.1\% for $\nu_\mu \to \nu_e$ and $\bar \nu_\mu \to \bar \nu_e$,
respectively. The corresponding beam and detector background has been
computed making use of the results quoted in
Ref.~\cite{Bouchez:2003fy,Gomez-Cadenas:2001eu}.

As far as the signal and background systematic errors, we followed the
arguments given in Ref.~\cite{jhf}, but we have considered only 2\%
and 5\%, and finally presented results for the worst case. The two
cases are on the other hand considered when sensitivity plots are
presented.

The event rates are given in Table~\ref{tab:spl:rates}. Note that,
given the difference in cross-section between neutrino and
antineutrino and that it is not possible to run a Super-Beam with both
polarities at the same time, to get comparable statistics the run
should be asymmetric in time: in Ref.~\cite{Bouchez:2003fy}, 2 years
run with main component $\nu_\mu$ and 8 years run with main component
$\bar{\nu}_\mu$ were assumed. It is however not clear that having a
comparable statistics is really necessary to get a good
$\theta_{13},\delta$ signal. In the case of the Neutrino Factory, for
example, it has been shown that a combination of two different
detectors, the first looking for $\nu_e \to \nu_\mu$ oscillations with
high statistics and the second for $\nu_e \to \nu_\tau$ with low
statistics is extremely useful to solve some of the parameter space
degeneration~\cite{silver}.  Indeed, with a 2+8 run the gain in the
$\bar \nu_\mu$ flux is compensated by a loss in the $\nu_\mu$ flux. We
have therefore also run in a symmetric 5+5 years configuration: our
results indicate that the two choices work similarly on the average,
with one or the other performing slightly better depending on the
particular region of the ($\theta_{13},\delta$) parameter space. In
the rest of the paper, to establish direct comparison
with~\cite{Bouchez:2003fy}, we adopt the 2+8
configuration.

\begin{table}[hbtp]
\begin{center}
\begin{tabular}{|c|c|c|c|c|c|c|} \hline
 $\theta_{13}$ & $\delta$ &  $ s_{atm}$ & $N_\nu$ &
  $N_{\bar{\nu}}$ & $P_{\nu_\mu \nu_e}(E=0.27)$ &
  $P_{\bar{\nu}_\mu \bar{\nu}_e}(E=0.25)$  \\ \hline \hline
 No Osc. &  &  & 24245 &  25467 &  & \\ \hline
   10 & 0 & + & 1200 &  1013 & 6.44$\times 10^{-2}$ & 5.68$\times 10^{-2}$ \\ \hline
   10 & 0 & - & 1033 &  1089 & 5.78$\times 10^{-2}$ & 6.52$\times 10^{-2}$ \\ \hline
   1 & 90 & + & 2 &  52 & 2.11$\times 10^{-5}$ & 3.13$\times 10^{-3}$ \\ \hline
   1 & 90 & - & 3 &  54 & 3.20$\times 10^{-5}$ & 3.27$\times 10^{-3}$ \\ \hline
   1 & -90 & + & 50 &  5 &  3.01$\times 10^{-3}$ & 6.96$\times 10^{-5}$ \\ \hline
   1 & -90 & - & 49 &  5 &  2.89$\times 10^{-3}$ & 5.16$\times 10^{-5}$ \\ \hline
\hline \hline
Beam back.        & &     &  92      &  110   & &  \\
\hline
Detector back.   & &     &   24     &  56 & &  \\
\hline \hline
\end{tabular}
\caption{\label{tab:spl:rates} \it Event rates for an exposure at a
standard Super-Beam. The oscillated charged-current events for
different values of $\theta_{13},\delta$ and sign of the atmospheric
mass difference, $s_{atm}$ for both neutrinos (2 years data taking)
and antineutrinos (8 years of data taking) are given. For comparison
with literature we show here the values obtained with the reference
parameters but $\theta_{23}=45^\circ$. The oscillation probabilities
at the mean neutrino/antineutrino energy (in GeV) are also shown.}
\end{center}
\end{table}


\subsection{Extraction of neutrino oscillation parameters in presence of signal}
\label{sec:sens}

The sensitivity to the $\theta_{13}$ and $\delta$ parameters has been
evaluated by assuming the signal and background rates reported in the
previous Sections and the following input values:
\mbox{$\Delta m_{12}^2 =7.3 \times 10^{-5} {\rm eV}^2$},
\mbox{$\theta_{12}=35^\circ$}, 
\mbox{$\Delta m_{23}^2 =  2.5 \times 10^{-3} {\rm eV}^2$},
and \mbox{$\theta_{23} = 40^\circ$}.
Since the sign of $\Delta m^2_{23}$ and the $\theta_{23}$-octant are
unknown, fits to both sign[$\Delta m^2_{23}$] = $\pm$ 1 and
sign[$\tan (2 \theta_{23})$] = $\pm$ 1 have been performed. In the
particular case $\theta_{23} = 45^\circ$, four out of eight solutions
of the systems of Eqs.~(\ref{eq:ene0int})-(\ref{eq:ene0t23sign})
disappear and only fits to the wrong assignment of sign($\Delta
m^2_{23}$) must be performed. However, since at a given confidence
level the contours for the allowed regions (around the theoretical
location of the true solution, the intrinsic clone and of the two sign
clones) are not qualitatively different from those we get for
$\theta_{23} = 40^\circ$, we have opted to present results for this
last case, only. 

\begin{figure}[t!]
\vspace{-0.5cm}
\begin{center}
\begin{tabular}{c}
\hspace{-0.3cm} \epsfxsize10cm\epsffile{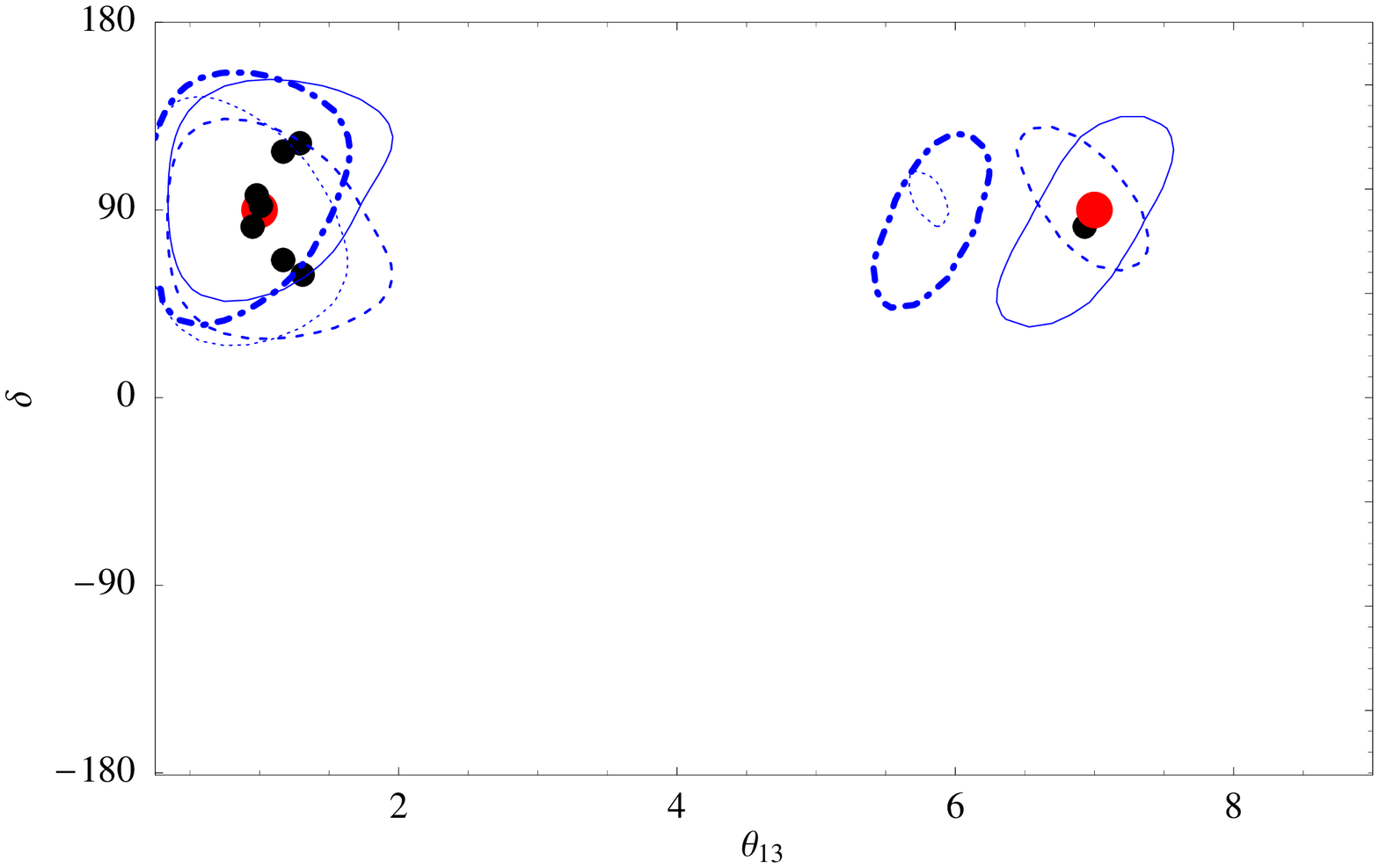} \\
\hspace{-0.3cm} \epsfxsize10cm\epsffile{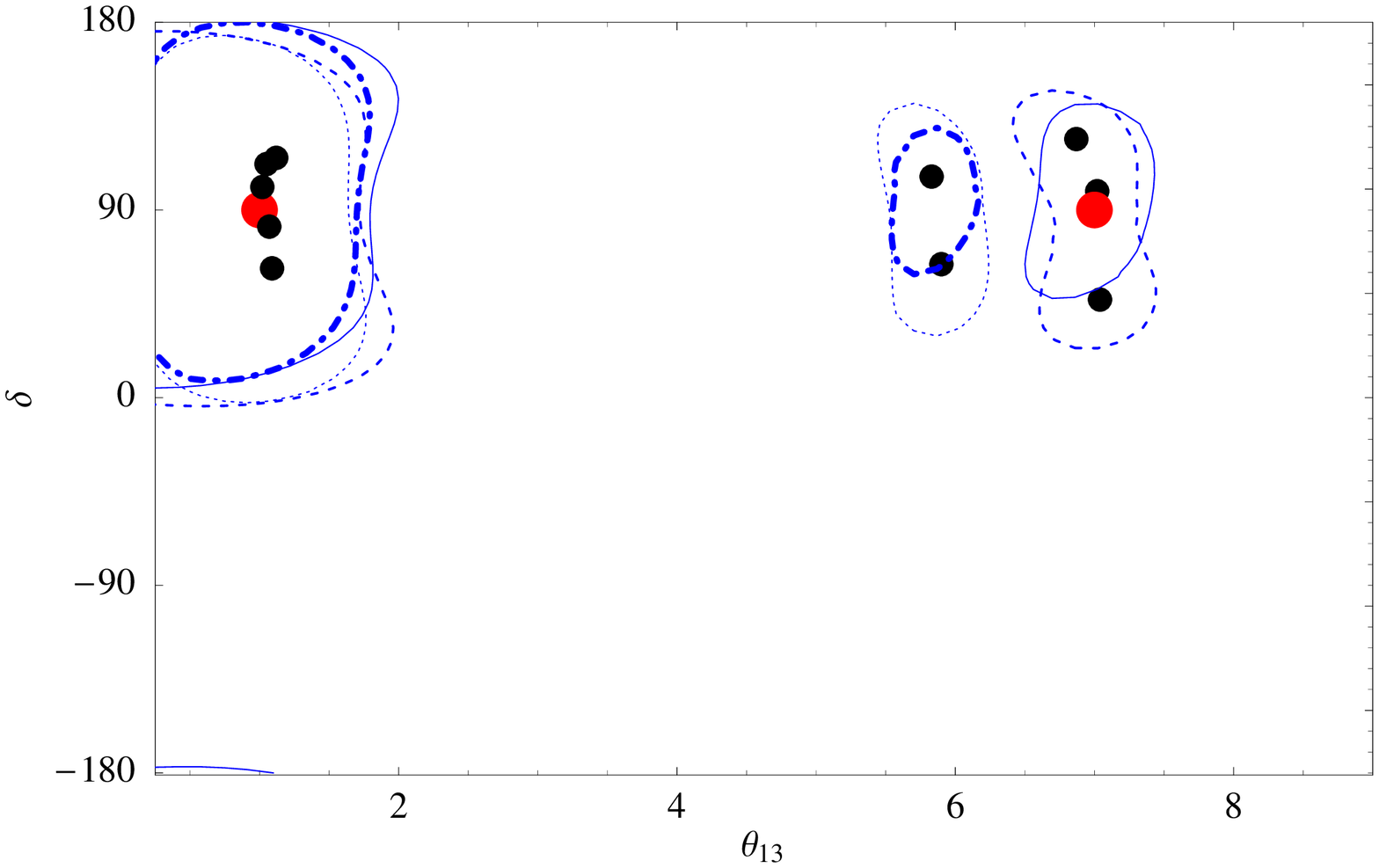} \\
\hspace{-0.3cm} \epsfxsize10cm\epsffile{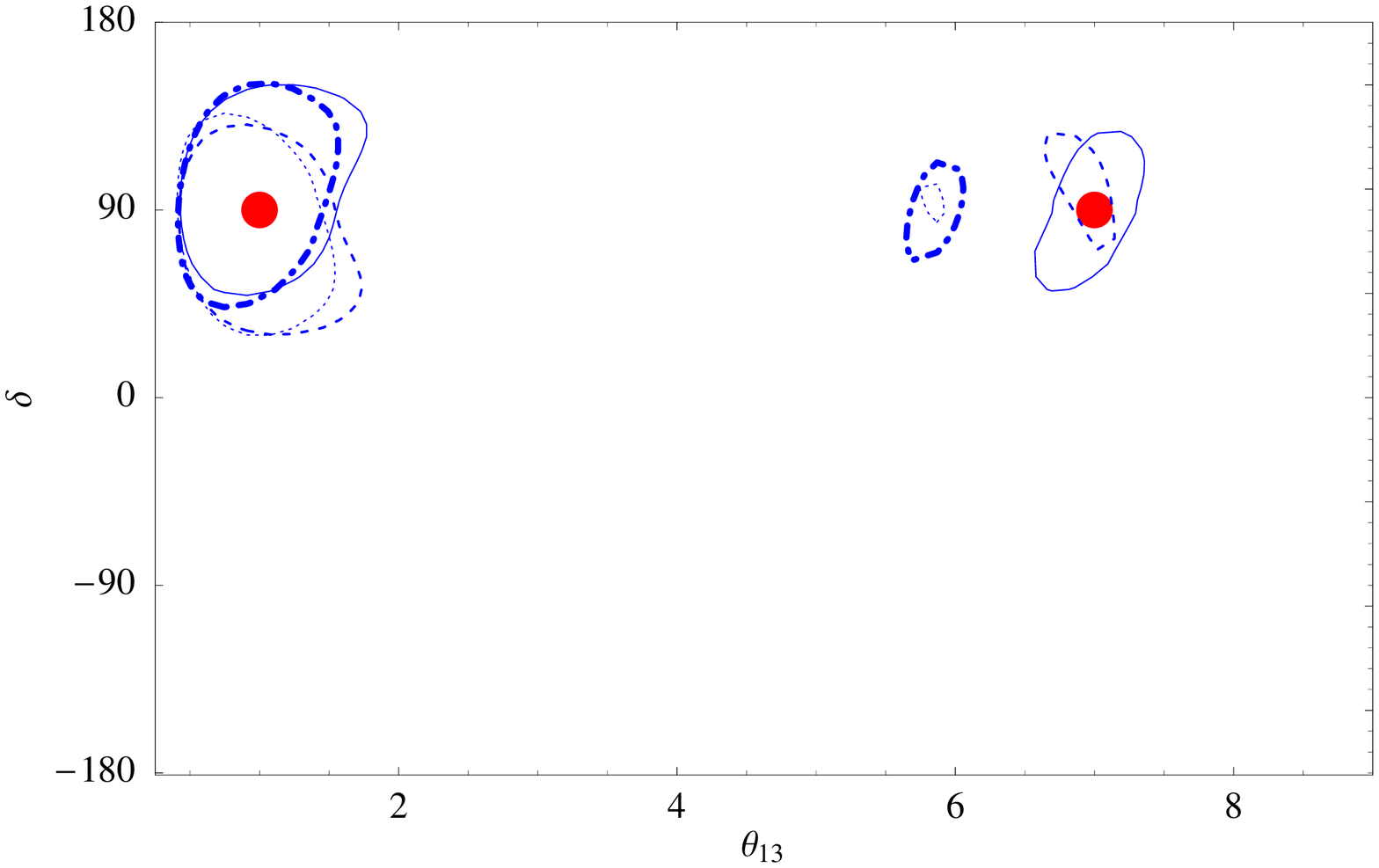} \\
\end{tabular}
\caption{\it Fits to $\theta_{13}$ and $\delta$ after a 10 yrs
$\beta$-Beam run and a 2+8 Super-Beam run. The 90\% CL contours are
shown for the following input values: $\theta_{13}=1^\circ,7^\circ$
and $\delta=90^\circ$. Upper panel: $\beta$-Beam results; middle panel:
Super-Beam results; lower panel combined results. Continuous lines stand 
for the intrinsic degeneracy; dashed lines stand for the sign degeneracy;
dot-dashed lines stand for the octant degeneracy; dotted lines stand for 
the mixed degeneracy. The light circle (red) shows the input value. Dark 
(black) dots are the theoretical clone locations computed as in 
Ref.~\cite{Donini:2003vz}.}
\label{fig:bb-sb-90}
\end{center}
\end{figure}
\begin{figure}[t!]
\vspace{-0.5cm}
\begin{center}
\begin{tabular}{c}
\hspace{-0.3cm} \epsfxsize10cm\epsffile{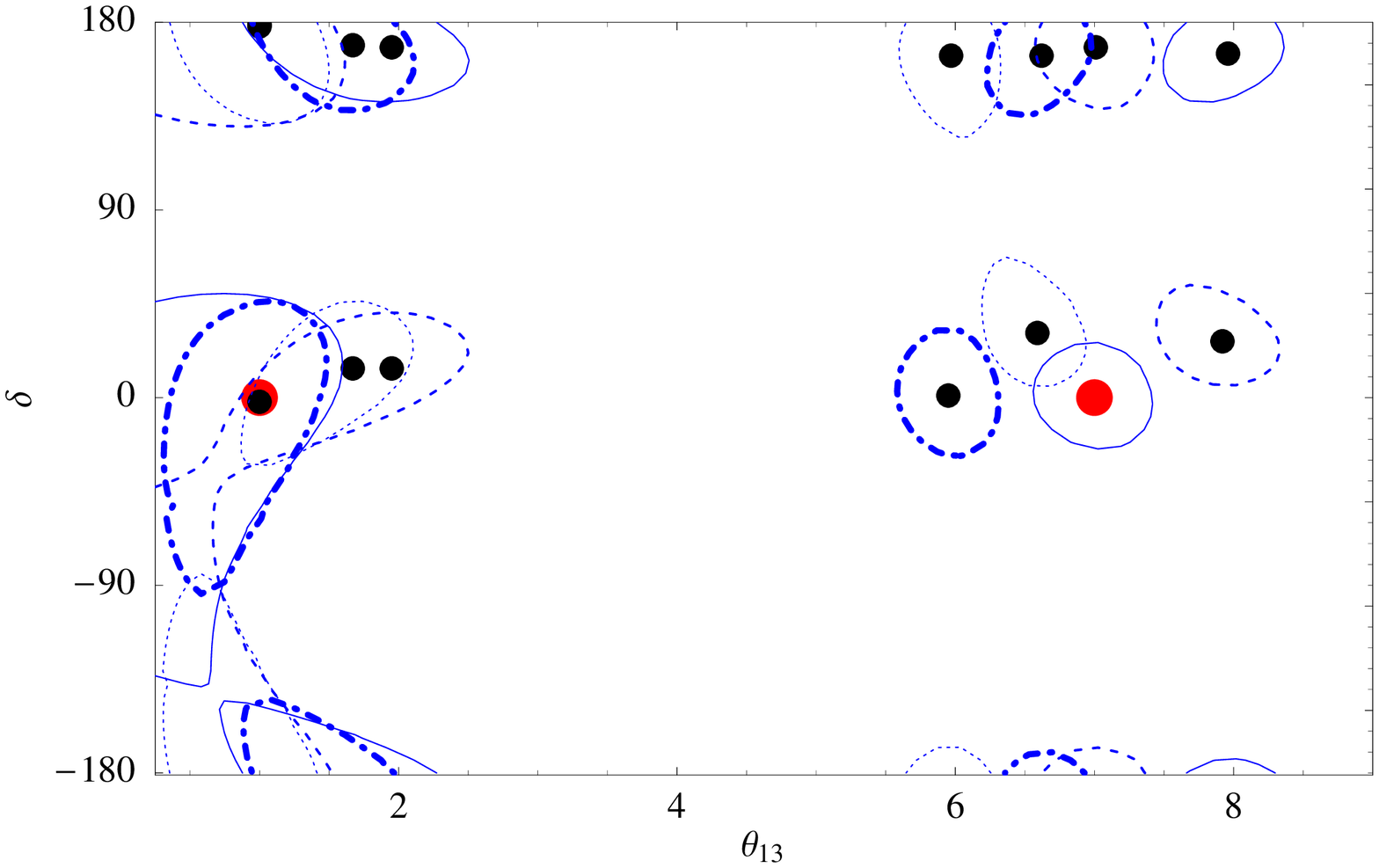} \\
\hspace{-0.3cm} \epsfxsize10cm\epsffile{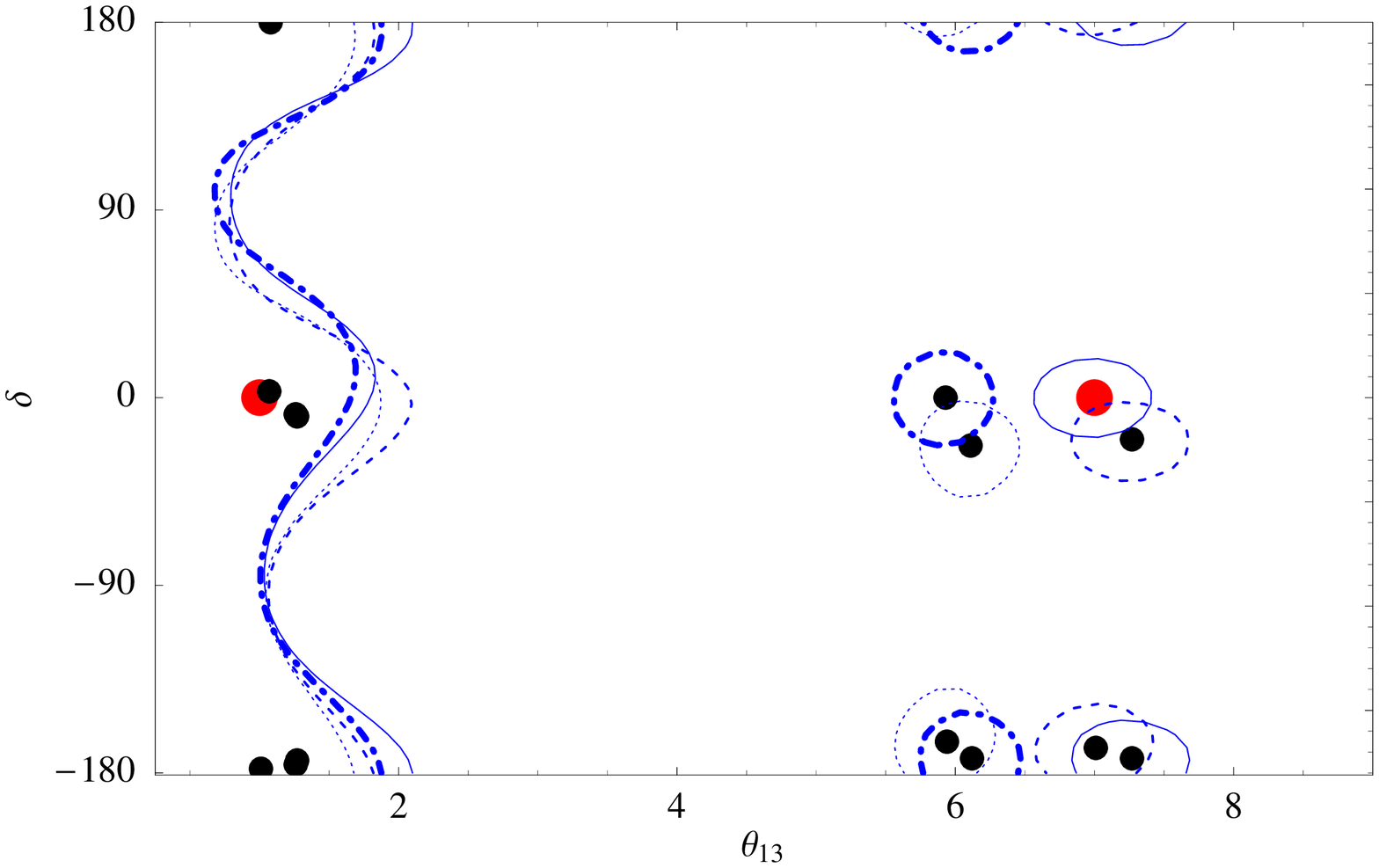} \\
\hspace{-0.3cm} \epsfxsize10cm\epsffile{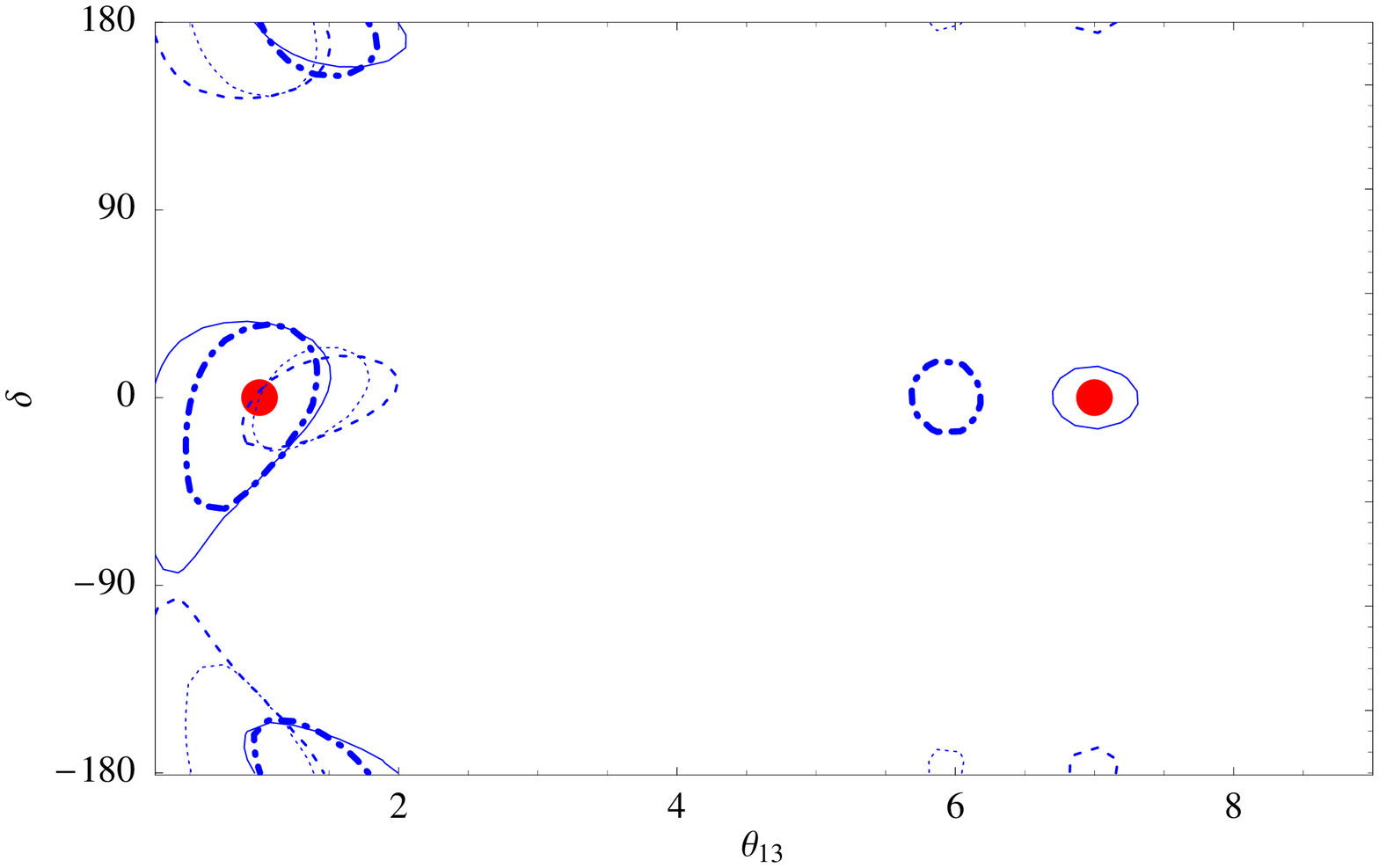} \\
\end{tabular}
\caption{\it Fits to $\theta_{13}$ and $\delta$ after a 10 yrs
$\beta$-Beam run and a 2+8 Super-Beam run. The 90\% CL contours are
shown for the following input values: $\theta_{13}=1^\circ,7^\circ$
and $\delta=0$. Upper panel: $\beta$-Beam results; middle panel:
Super-Beam results; lower panel: combined results. Continuous lines
stand for the intrinsic degeneracy; dashed lines stand for the sign
degeneracy; dot-dashed lines stand for the octant degeneracy; dotted
lines stand for the mixed degeneracy. The light circle (red) shows the
input value. Dark (black) dots are the theoretical clone locations 
computed as in Ref.~\cite{Donini:2003vz}.}
\label{fig:bb-sb-00}
\end{center}
\end{figure}
\begin{figure}[t!]
\vspace{-0.5cm}
\begin{center}
\begin{tabular}{c}
\hspace{-0.3cm} \epsfxsize10cm\epsffile{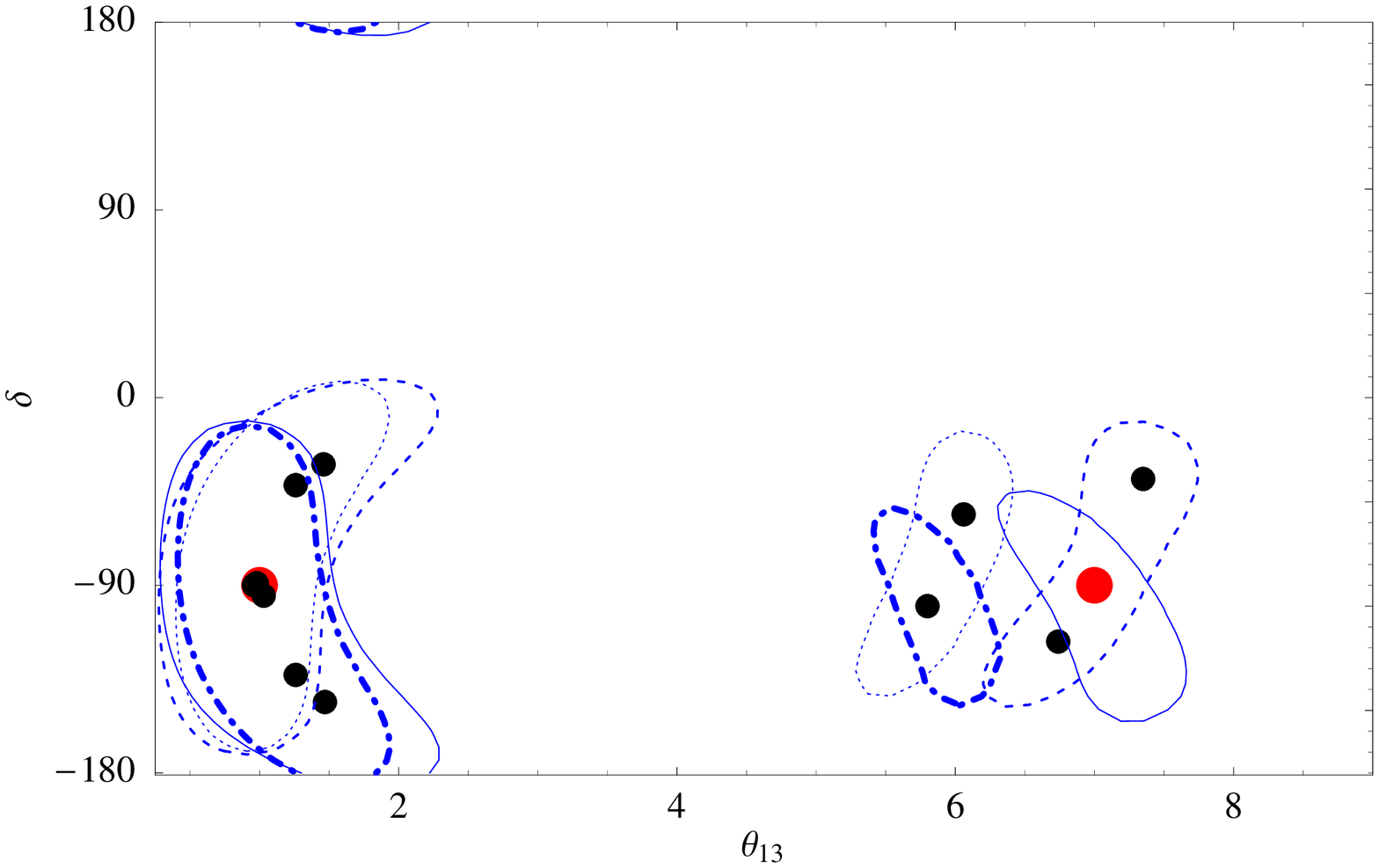} \\
\hspace{-0.3cm} \epsfxsize10cm\epsffile{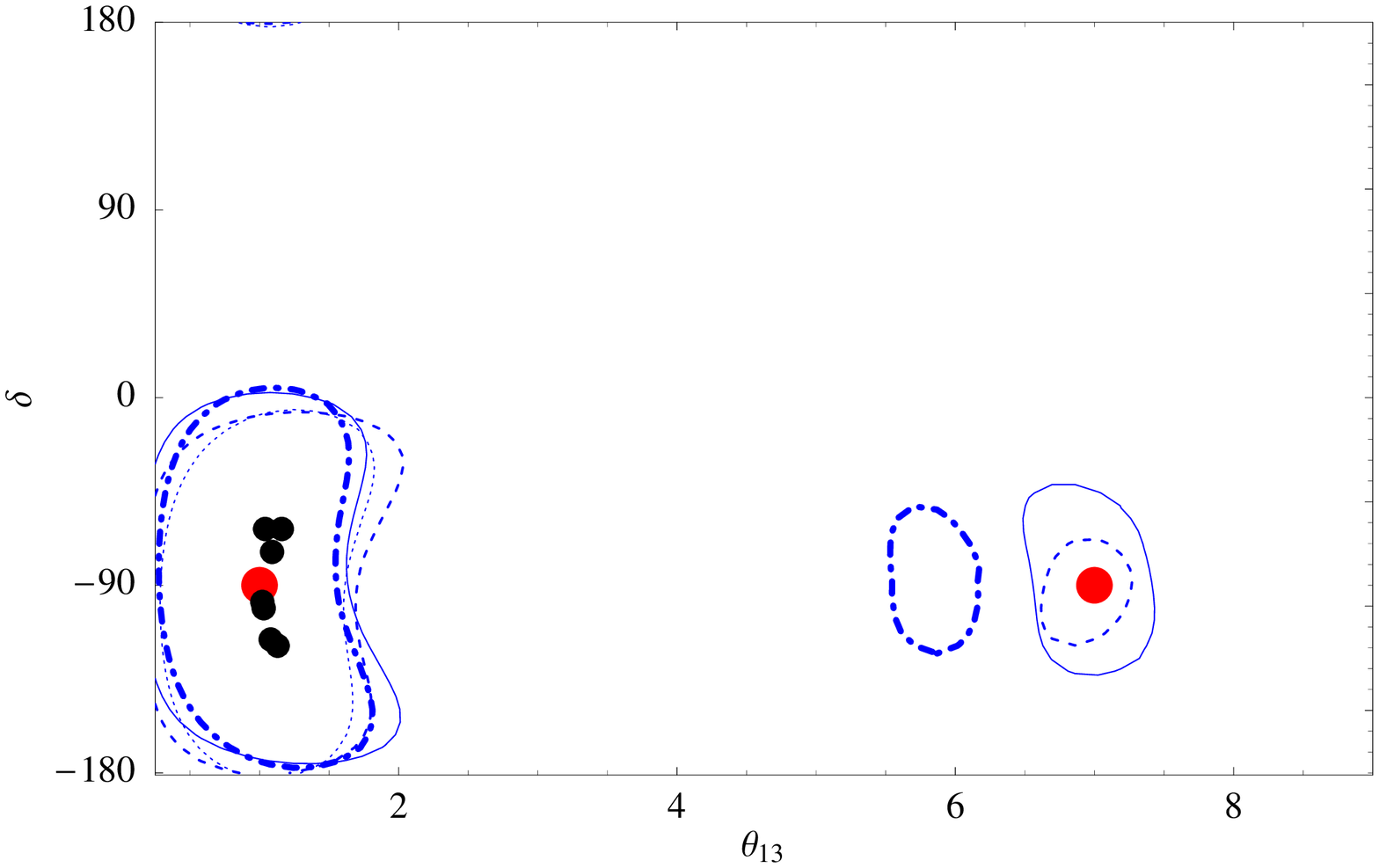} \\
\hspace{-0.3cm} \epsfxsize10cm\epsffile{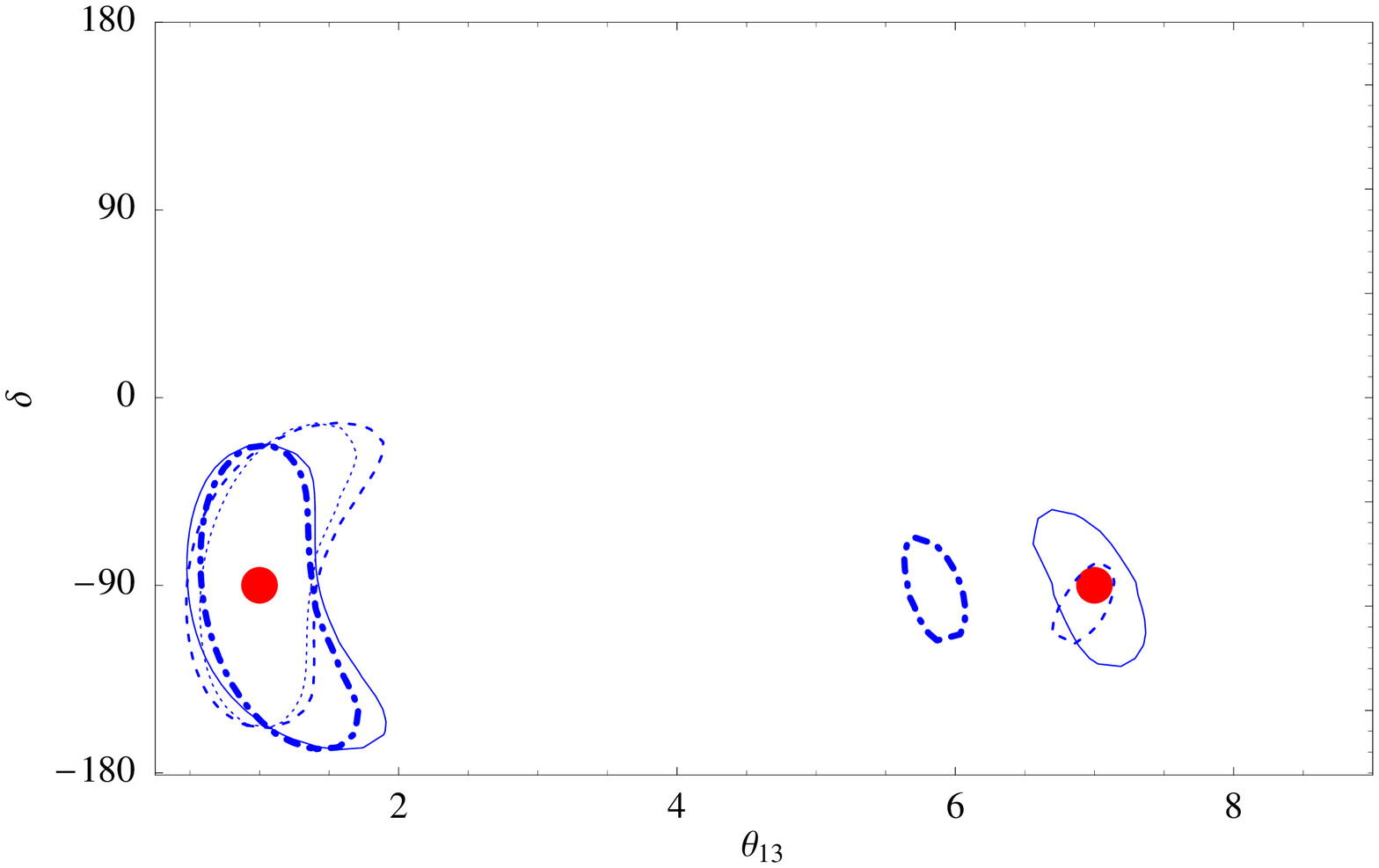} \\
\end{tabular}
\caption{\it Fits to $\theta_{13}$ and $\delta$ after a 10 yrs
$\beta$-Beam run and a 2+8 Super-Beam run. The 90\% CL contours are
shown for the following input values: $\theta_{13}=1^\circ,7^\circ$
and $\delta=-90^\circ$. Upper panel: $\beta$-Beam results; middle panel:
Super-Beam results; lower panel: combined results. Continuous lines
stand for the intrinsic degeneracy; dashed lines stand for the sign
degeneracy; dot-dashed lines stand for the octant degeneracy; dotted
lines stand for the mixed degeneracy. The light circle (red) shows the
input value. Dark (black) dots are the theoretical clone locations 
computed as in Ref.~\cite{Donini:2003vz}.}
\label{fig:bb-sb-M0}
\end{center}
\end{figure}

Our results for two specific values, $\bar \theta_{13} =
1^\circ,7^\circ$, are presented in
Figs.~\ref{fig:bb-sb-90}-\ref{fig:bb-sb-M0}. In each figure we plot
the 90 \% CL contours for the two considered values of $\bar
\theta_{13}$ and a fixed value of $\delta$: $\bar \delta = 90^\circ$
(Fig.~\ref{fig:bb-sb-90}), $\bar \delta = 0$ (Fig.~\ref{fig:bb-sb-00})
and $\bar \delta = - 90^\circ$ (Fig.~\ref{fig:bb-sb-M0}). From top to
bottom, the results for the $\beta$-Beam, for the Super-Beam and for
the combination of the two are presented. 


In every separate case, the four possible choices of the discrete
variables $s_{atm},s_{oct}$ are reported: continuous lines stand for
the true solution and the intrinsic degeneracy (right sign($\Delta
m^2_{23}$) and right $\theta_{23}$-octant); dashed lines stand for the
sign degeneracy (wrong sign($\Delta m^2_{23}$) and right
$\theta_{23}$-octant); dot-dashed lines stand for the octant
degeneracy (right sign($\Delta m^2_{23}$) and wrong
$\theta_{23}$-octant); dotted lines stand for the mixed degeneracy
(wrong sign($\Delta m^2_{23}$) and wrong $\theta_{23}$-octant). In all
the figures, the light (red) circle shows the input value. The dark
(black) circles in the plots for the $\beta$-Beam and the Super-Beam
represent the theoretical clone locations computed as in
Ref.~\cite{Donini:2003vz} for the two facilities considered in this
paper. They have been included to show the rather good agreement
between the theoretical computation and the output of the fits. Notice
that for $\bar \theta_{13} = 7^\circ$ some of the theoretical clones
are missing. Indeed, an analytic solution of the systems in
Eqs. (\ref{eq:ene0int})-(\ref{eq:ene0t23sign}) is not always found for
large $\bar \theta_{13}$ values, see~\cite{Donini:2003vz} for
details.

Notice first that the impact of the octant-ambiguity in all the
considered cases is by far more relevant for large values of $\bar
\theta_{13}$ than for small values. For the three values of $\delta$
that we have analyzed we get the octant clones with a $\Delta
\theta_{13} \sim 1^\circ$ shift with respect to the true solution when
$\bar \theta_{13} = 7^\circ$, whereas we get a significantly smaller
shift when $\bar \theta_{13} = 1^\circ$. This comment apply both to
the $\beta$-Beam and the Super-Beam results: indeed, when the
octant-ambiguity is considered, both facilities show pretty similar
contours. This is why, as a general result, the combination of the two
facilities does not solve this ambiguity. Even in the case of
$\bar\theta_{13} = 7^\circ$, for which a significant statistics is
accumulated for both facilities, the bottom plot of
Figs.~\ref{fig:bb-sb-90}-\ref{fig:bb-sb-M0} shows that at 90 \% CL the
true solution and the octant ambiguity survive for the three values of
$\delta$ considered.

Concerning the sign ambiguity we notice that, for both facilities and
for both considered values of $\bar \theta_{13}=1^\circ,7^\circ$, the
allowed regions corresponding to the clone solutions overlap with the
true solution in a significant way. As a consequence the combination
of the two facilities, albeit reducing the 90 \% CL contours, does not
solve the sign ambiguity, either. However, contrary to the case of the
octant degeneracy treated above, this does not affect the measurement
of the two continuous unknowns $\theta_{13}$ and $\delta$. For large
$\bar\theta_{13}$, a considerably good measurement of both quantities
is achieved, with no clue on the mass hierarchy. An exception to this
statement is the particular case $\bar \theta_{13} = 7^\circ, \bar
\delta = 0$, for which the sign clones are quite definite regions
located in different places for the $\beta$-Beam and the Super-Beam.
This actually allows the cancellation of some of the allowed regions
when combining the two facilities. The ambiguities are however not
solved, since clone regions survive near $\delta = 180^\circ$.

Discussing the mixed ambiguity, we notice that much of what was said
for the sign degeneracies can be repeated in this case: the only
difference being that allowed regions around mixed clones overlap
significantly with the octant degeneracy, and not with the true
solution as it was the case for the sign degeneracy. As a consequence,
mixed clones survive after the combination of the two facilities. This
comment applies to all the considered input pairs, with the exception
of the $\bar \theta_{13} = 7^\circ, \bar \delta = 0$ case (as for the
sign ambiguity) and of $\bar \theta_{13} = 7^\circ,\bar \delta =
-90^\circ$, for which no mixed ambiguity is present in the Super-Beam.

We observe that in general four allowed regions are still present
after adding data from the $\beta$-Beam and the Super-Beam at large
$\bar \theta_{13}$ (with the exception $\bar \theta_{13} =
7^\circ,\bar \delta = -90^\circ$, for which the mixed clone is absent
after combination). These four regions either overlap in pairs (case
of $\bar \delta = 90^\circ, -90^\circ$) or stay well apart (case of
$\bar \delta = 0$), in all cases allowing a measure of the two
continuous parameters but not of the two discrete parameters. On the
other hand, for small $\bar \theta_{13}$ we are clearly on the
sensitivity limit of these experiments: no clear measure of $\delta$
is possible (at 90 \% CL we can just state if $\delta$ is positive or
negative) and no discrete ambiguity is solved.

From the results reported in the top plots of
Figs.~\ref{fig:bb-sb-90}-\ref{fig:bb-sb-M0}, it is clear that by using
the $\beta$-Beam or the Super-Beam as isolated experiment it is not
possible to solve any of the degeneracies, although for large enough
$\bar \theta_{13}$ a first estimate of the two continuous parameters
$\theta_{13}$ and $\delta$ can be attempted. Even when combining the
two experiments we have found that as a general result the discrete
parameters are not measured. These results are in contrast with the
statement of~\cite{Bouchez:2003fy} {\it `` We stress the fact that an
experiment working at very short baselines has the smallest possible
parameter degeneracies and ambiguities and it is the cleanest possible
environment where to look for genuine leptonic CP violation
effects''}. The correct statement is that in many cases the discrete
ambiguities, although not solved, does not affect in a significant way
the measure of the continuous parameters. Notice, however, that in
general multiple solutions are found with either larger uncertainties
in both parameters when these regions overlap or a proliferation of
disconnected regions in the parameter space\footnote{The presence and
the locations of the clones as derived in this paper are in rather
good agreement with the analytical calculations quoted
in~\cite{Donini:2003vz}.}.


%

%

\subsection{Exclusion plots in absence of signal}
\label{sec:sensi}

\begin{figure}[t!]
\vspace{-0.5cm}
\begin{center}
\begin{tabular}{c}
\hspace{-0.3cm} \epsfxsize10cm\epsffile{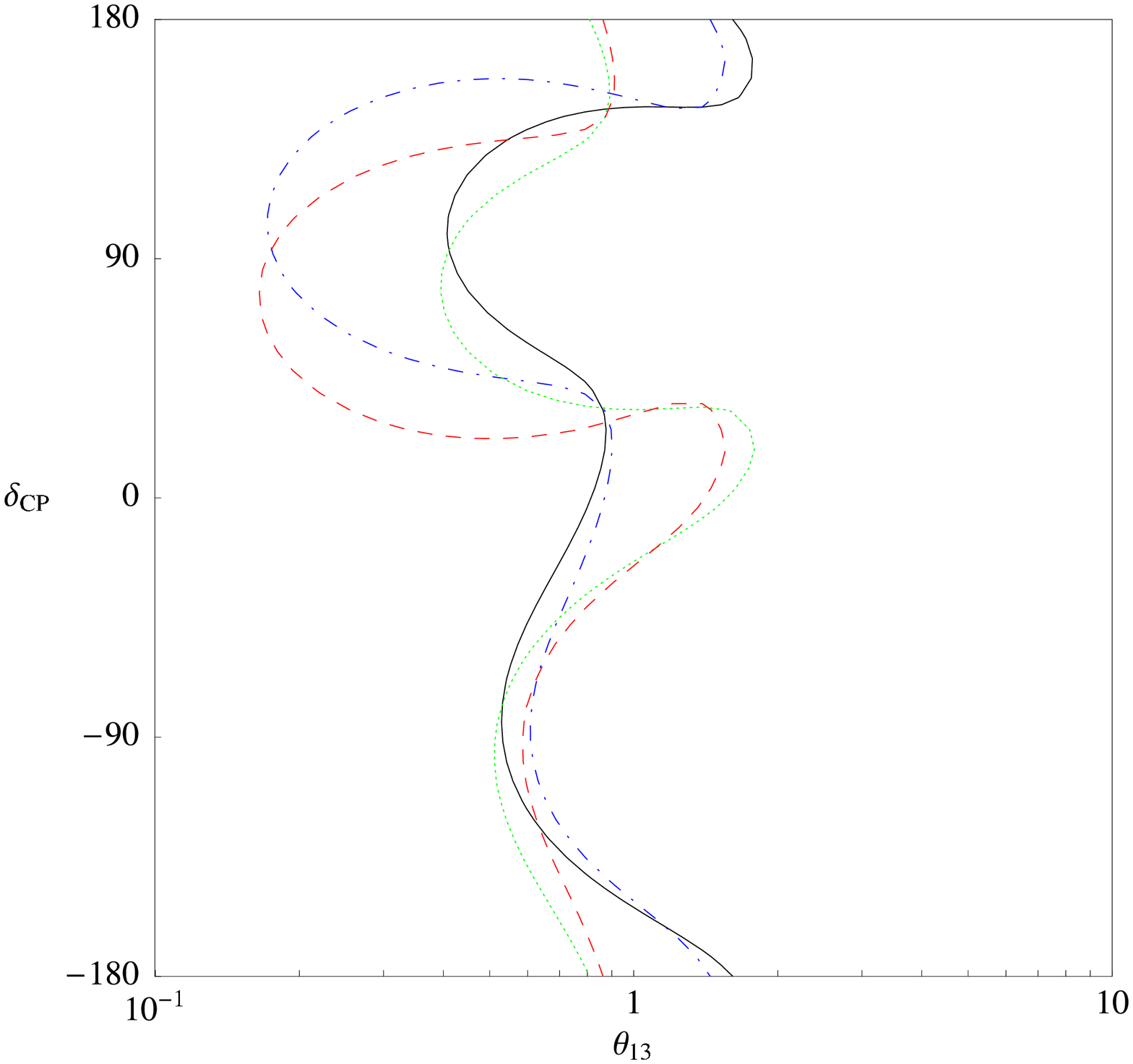}\\
\hspace{-0.3cm} \epsfxsize10cm\epsffile{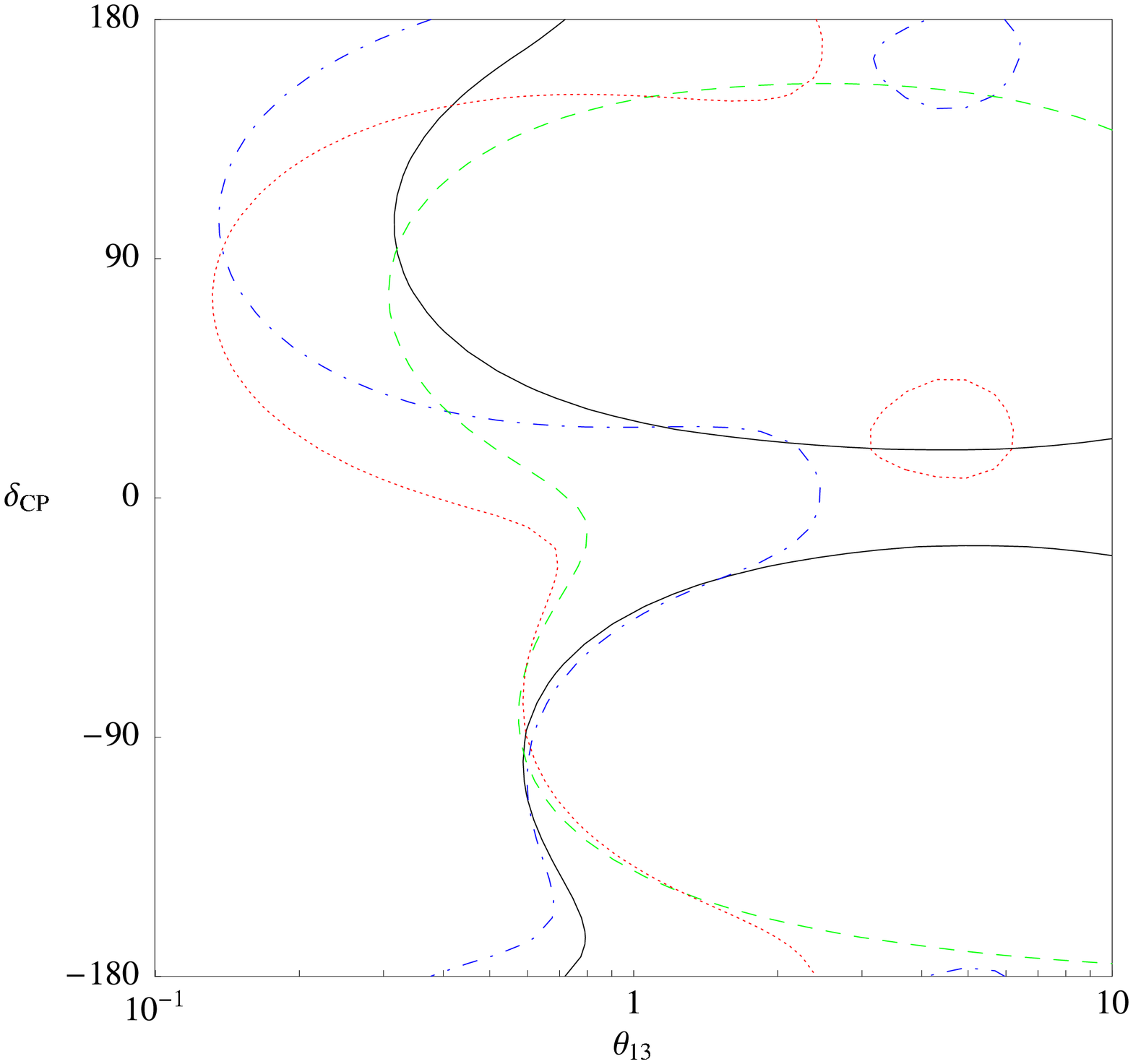}
\end{tabular}
\caption{\it 
Sensitivity plots to $\theta_{13}$ (upper) and $\delta$ (lower) after 
a 10 yrs $\beta$-Beam run. The 90\% CL contours are shown. Continuous 
lines stand for the intrinsic degeneracy; dashed lines stand for the sign
degeneracy; dot-dashed lines stand for the octant degeneracy; dotted lines 
stand for the mixed degeneracy.}
\label{fig:sensiBB}
\end{center}
\end{figure}

\begin{figure}[t!]
\vspace{-0.5cm}
\begin{center}
\begin{tabular}{c}
\hspace{-0.3cm} \epsfxsize10cm\epsffile{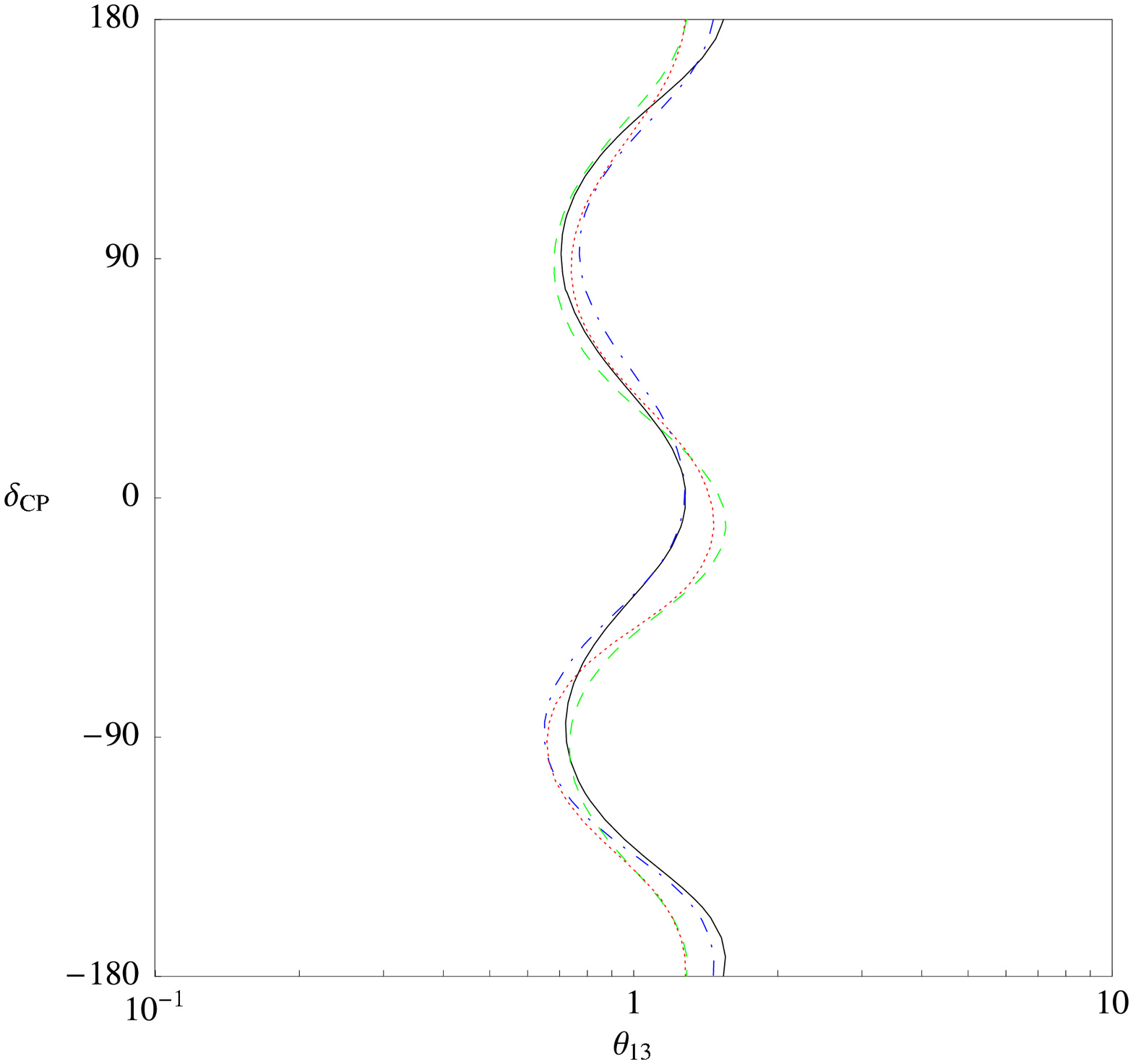}\\
\hspace{-0.3cm} \epsfxsize10cm\epsffile{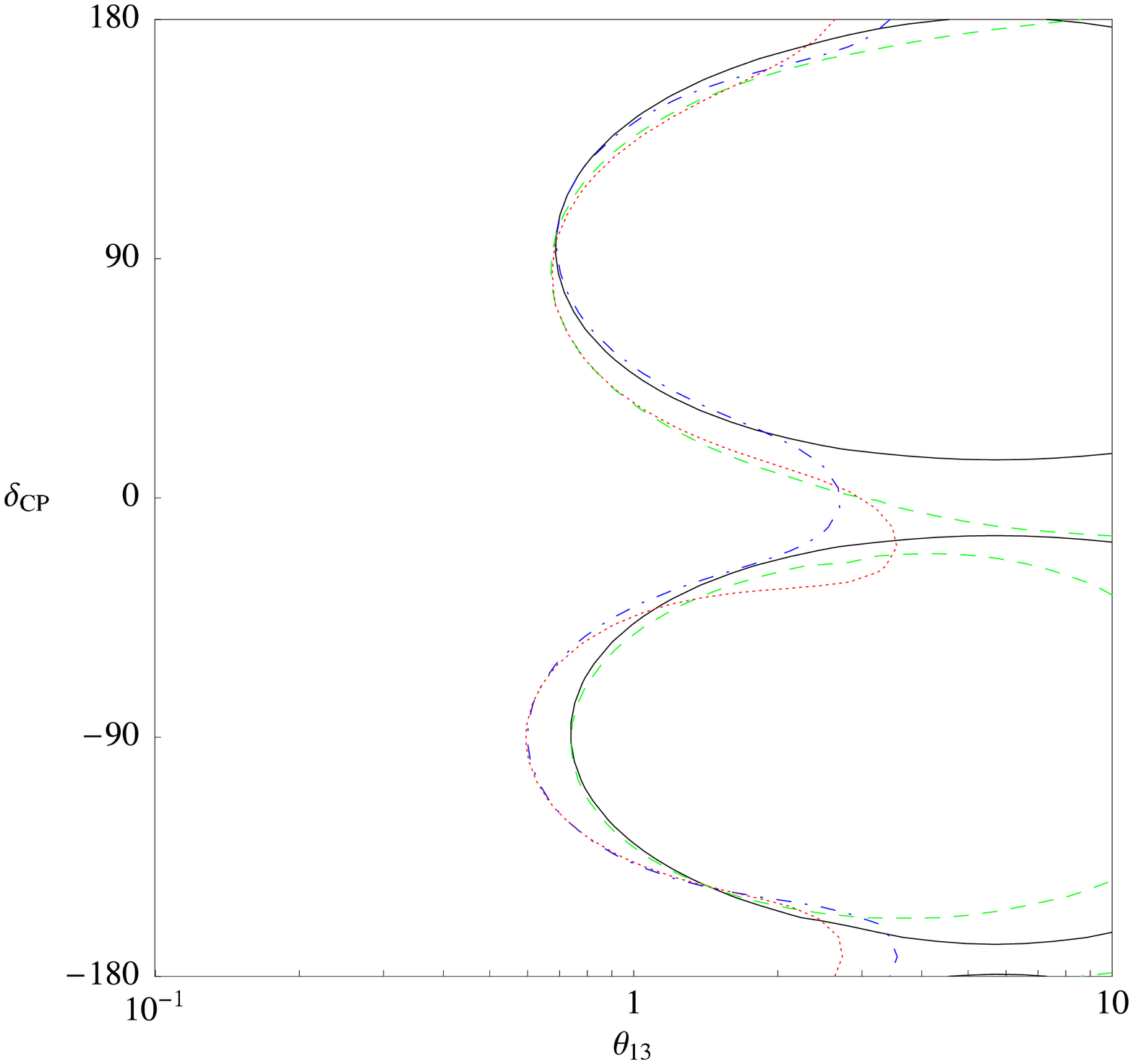}
\end{tabular}
\caption{\it 
Sensitivity plots to $\theta_{13}$ (upper) and $\delta$ (lower) after 
a 2+8 Super-Beam run. The 90\% CL contours are shown. Continuous lines 
stand for the intrinsic degeneracy; dashed lines stand for the sign
degeneracy; dot-dashed lines stand for the octant degeneracy; dotted 
lines stand for the mixed degeneracy.}
\label{fig:sensiSB}
\end{center}
\end{figure}

\begin{figure}[t!]
\vspace{-0.5cm}
\begin{center}
\begin{tabular}{c}
\hspace{-0.3cm} 
      \epsfxsize8cm\epsffile{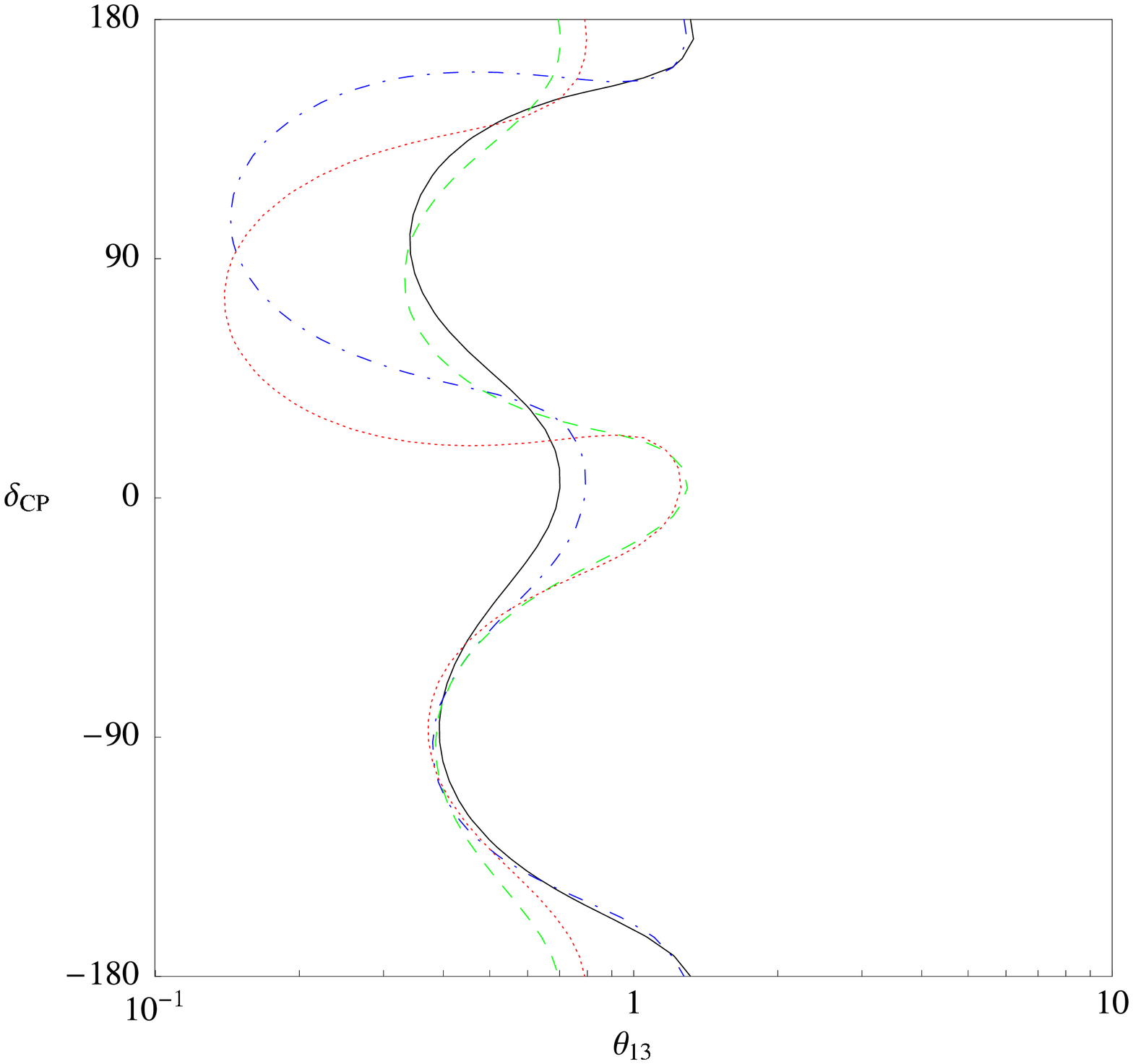} 
      \epsfxsize8cm\epsffile{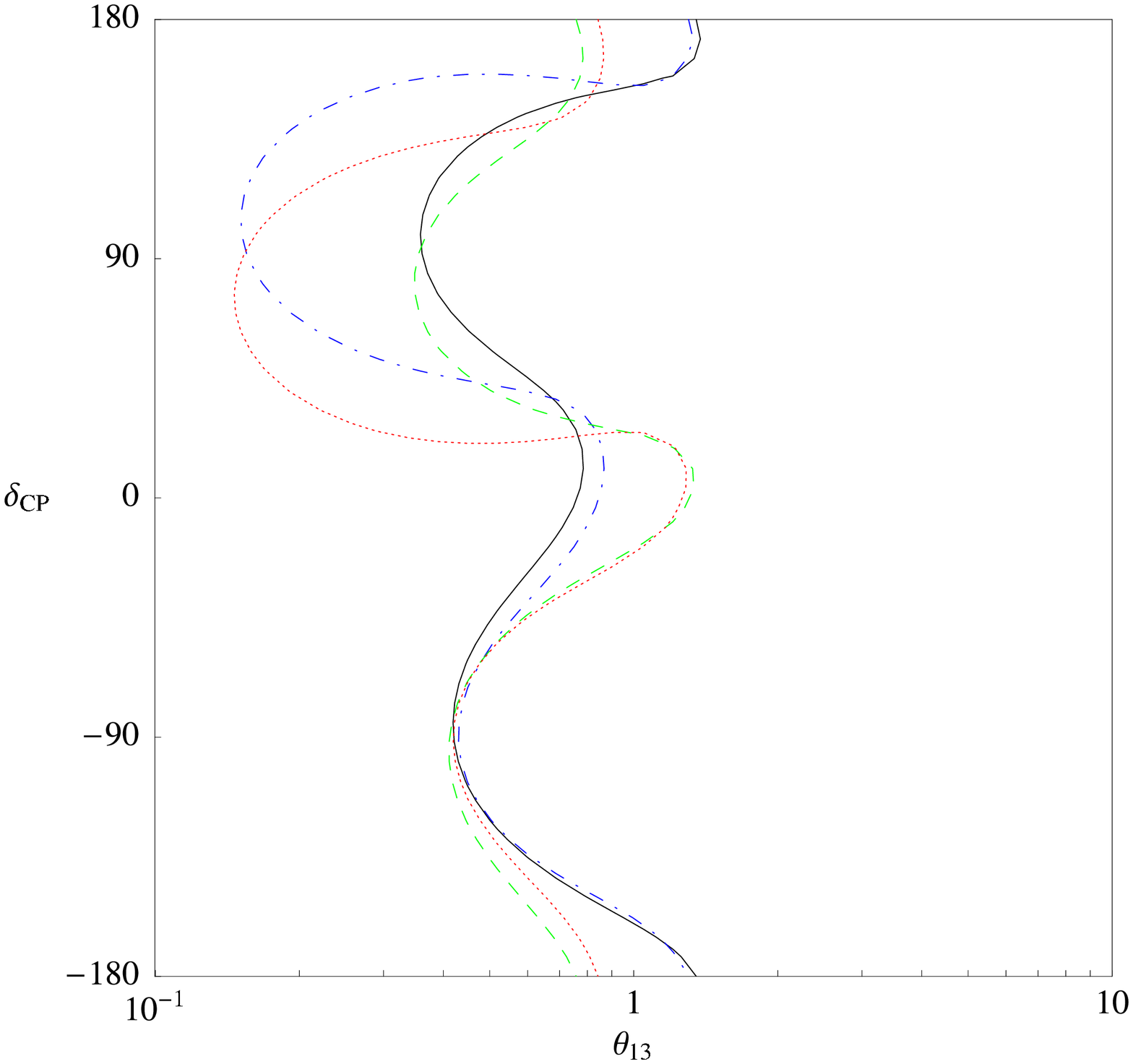} \\
\hspace{-0.3cm} 
      \epsfxsize8cm\epsffile{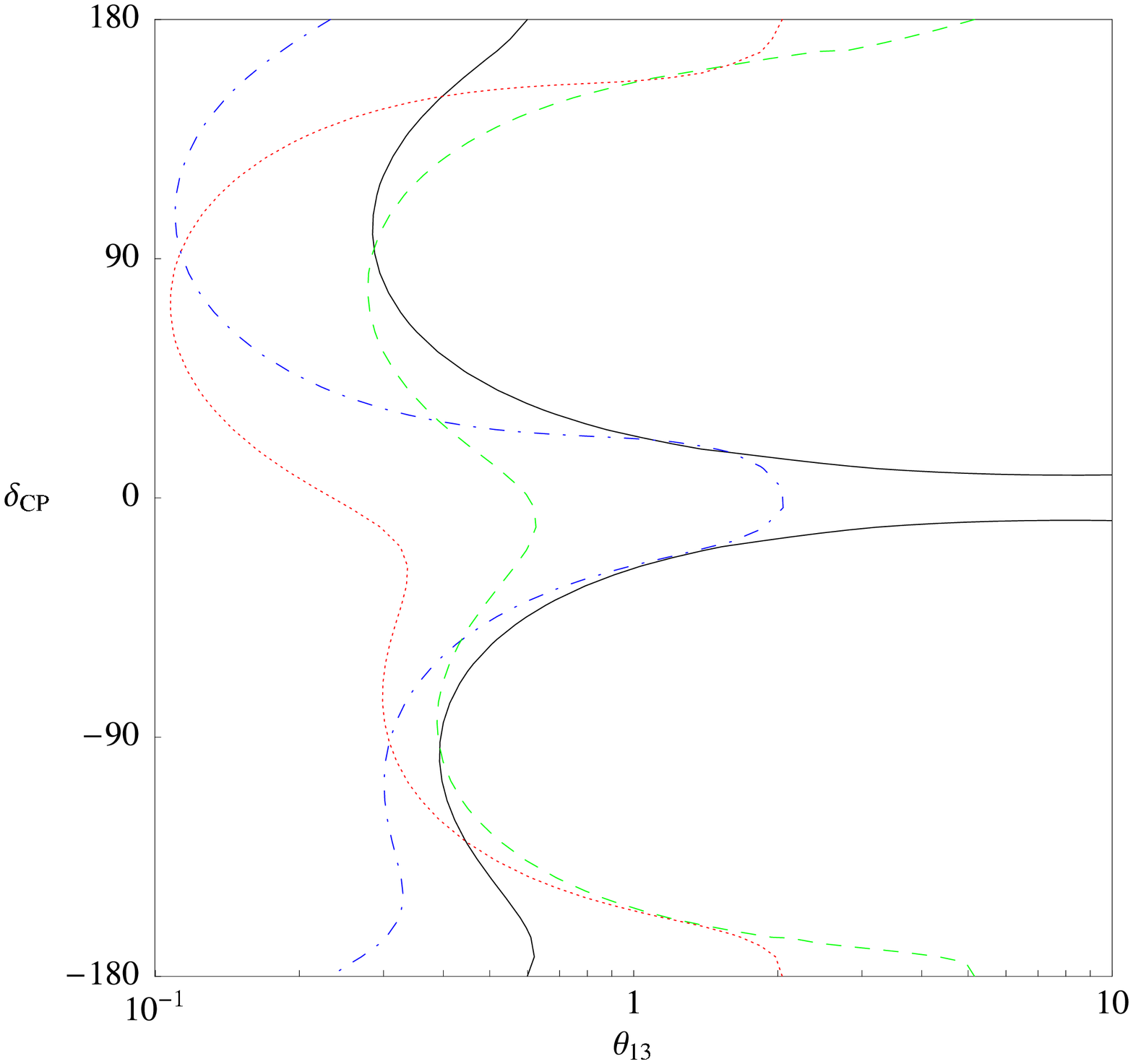}
      \epsfxsize8cm\epsffile{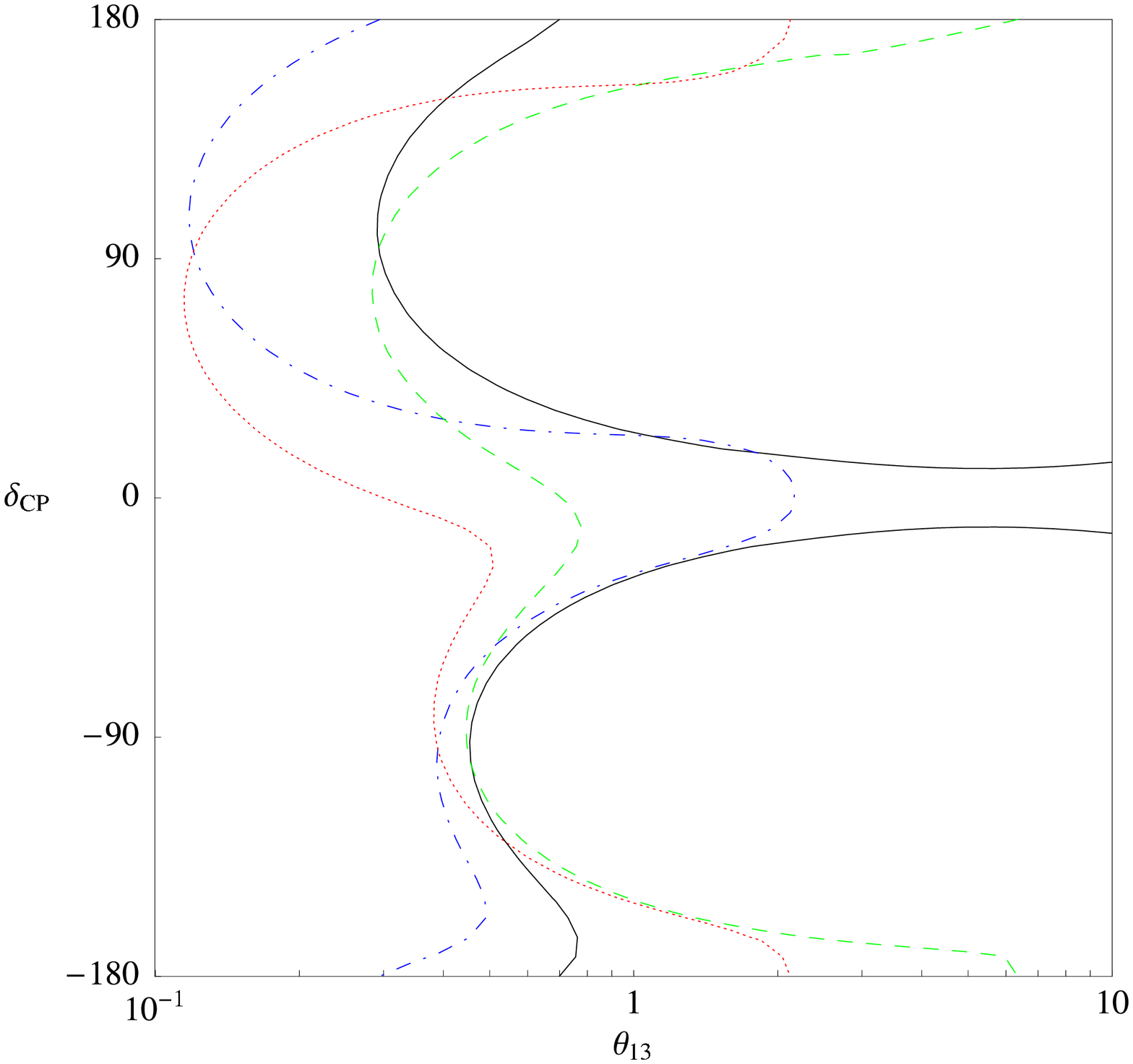}\\
\end{tabular}
\caption{\it 
Sensitivity plots to $\theta_{13}$ (upper) and $\delta$ (lower) after a 
combined 10 yrs $\beta$-Beam and a 2+8 Super-Beam run. The 90\% CL contours 
for two values of the systematic errors are shown: 2\% (left panels) and 
5\% (right panels). Continuous lines stand for the intrinsic degeneracy; 
dashed lines stand for the sign degeneracy; dot-dashed lines stand for the 
octant degeneracy; dotted lines stand for the mixed degeneracy.}
\label{fig:sensiBBSB2}
\end{center}
\end{figure}

In Figs.~\ref{fig:sensiBB},~\ref{fig:sensiSB},~\ref{fig:sensiBBSB2} we
present exclusion plots in the absence of a signal for the
$\beta$-Beam, the SPL-Super-Beam and their combination. The different
lines represent, with the notation used in the previous figures,
different choices of the two discrete parameters $s_{atm},s_{oct}$.

In the upper plot of
Fig.~\ref{fig:sensiBB},~\ref{fig:sensiSB},~\ref{fig:sensiBBSB2} we
draw the 90\% CL contour defining the sensitivity limit on
$\theta_{13}$ in case of absence of a signal, with $\delta$ as a free
parameter. Notice how the sensitivity limit spans a region from
$\sim0.5^\circ$ to $2.5^\circ$, the less stringent limit being for
$\delta = 0$. A significant loss in sensitivity for this value of
$\delta$ is induced by the unsolved discrete ambiguities, whereas for
other $\delta$ values the worse limit is generally given by the true
solution. It is worth noting that the $\beta$-Beam plots of
Fig.~\ref{fig:sensiBB} show a better sensitivity for $\delta =
90^\circ$ than for $\delta = -90^\circ$ (see also
\cite{Bouchez:2003fy}). This result reflects the expected large signal
statistics for neutrinos at $\delta=90^\circ$ and the small background
(1 event) for antineutrinos. Being the background affected by large
uncertainties (see Section~\ref{cross}), we checked that, if the
expected number of background events increases, the sensitivity
enhancement disappears and the plots become symmetric in $\delta$.  In
the case of the Super-Beam, Fig.~\ref{fig:sensiSB}, we notice no
asymmetry in $\delta$ in both plots. This is a consequence of the more
symmetric background reported in Tab. 3 for both the neutrino and
antineutrino beams. Furthermore, we point out that the effect of the
degeneracies on the sensitivity is much lower than in the case of the
$\beta$-Beam: in both plots it can be seen how the different excluded
regions coming from different choices of the discrete parameters
$s_{atm}$ and $s_{oct}$ overlap significantly.

In the lower plot of
Figs.~\ref{fig:sensiBB},~\ref{fig:sensiSB},~\ref{fig:sensiBBSB2} we
draw the 90\% CL contour in the absence of a CP violating signal (that
can be interpreted as $\delta= 0^\circ$) for fixed values of
$\theta_{13}$. In this case it is clearly visible how the true
solution define an upper bound on $|\delta| \leq 30^\circ-40^\circ$,
whereas the sign degeneracy draw a lower bound for $|\pi - \delta|
\geq 150^\circ - 170^\circ$. The octant and mixed degeneracy, on the
other hand, do not play a significant role in this sensitivity plot
apart for excluding small regions of parameter space around
$\theta_{13} \simeq 4^\circ$ and positive $\delta$.

We also studied the dependence of the sensitivity on the systematic
error for the combined $\beta$-Beam and Super-Beam setup. The results
are shown in Fig.~\ref{fig:sensiBBSB2}, from which it can be noticed that 
there is not a strong impact on the sensitivity going from 5\% to 2\%.

\section{Conclusion}
Over the past years the possibility to build a CERN neutrino complex,
based on the novel concept of the $\beta$-Beam and on a neutrino
Super-Beam, that exploits a 1 Megaton water Cerenkov detector located
at the Fr\'{e}jus underground laboratory (130 km baseline) has been
put forward. In this paper we study for the first time the eightfold
degeneracy for such a scenario.

After a brief theoretical discussion about the eightfold degeneracy
and a short description of the $\beta$-Beam and Super-Beam facilities,
we focus on the neutrino and antineutrino cross-sections at low
energies. At the time the neutrino complex will become operational, it
would be possible to measure with high accuracy the cross-sections.
However, nowadays we have the problem to compute the physics potential
of a facility having in mind that the expected number of signal and
background events strongly depend on the adopted calculation. In
particular we pointed out that the background in the antineutrino
channel of a $\beta$-Beam depends on the shape and absolute value of
the cross-section and that this background can significantly affect
the sensitivity to $\theta_{13}$ and $\delta$.

From the results reported in this paper it is clear that by using the
$\beta$-Beam or the Super-Beam alone it is not possible to solve all
the degeneracies, although for large enough $\bar \theta_{13}$ a first
estimate of the two continuous parameters $\theta_{13}$ and $\delta$
can be attempted. Even when combining the two experiments we have
found that as a general result the discrete parameters are not
measured. In many cases the discrete ambiguities, although not solved,
do not affect in a significant way the measurement of the continuous
parameters. Notice, however, that in general multiple solutions are
found with either larger uncertainties in both parameters
when these regions overlap or a proliferation of disconnected regions
in the parameter space.

As a final comment we want to stress that being the experiment
discussed in this paper a ``counting experiment'', the small
differences between the fluxes are averaged-out. Therefore, being
$N_{evts}(\nu_\mu\rightarrow\nu_e)\simeq
T(N_{evts}(\nu_e\rightarrow\nu_\mu))$, the combination (synergy) of a
$\beta$-Beam and of a Super-Beam only determines an increase of
statistics for both neutrinos and antineutrinos. Therefore, there is
not a real synergy.


\section*{Acknowledgments}

We would like to thank B.~Gavela, S.~Gilardoni, J.~Gomez-Cadenas,
P.~Hernandez, P.~Lipari, O.~Mena, M.~Mezzetto and F.~Terranova for
useful discussions during the preparation of this paper.
 


\begin{thebibliography}{999}

\bibitem{atmo}
Y.~Fukuda {\it et al.}  [Super-Kamiokande Collaboration],
Phys.\ Rev.\ Lett.\  {\bf 81} (1998) 1562 [arXiv:hep-ex/9807003];\\
M.~Ambrosio {\it et al.}  [MACRO Collaboration],
Phys.\ Lett.\ B {\bf 517} (2001) 59
[arXiv:hep-ex/0106049].

\bibitem{Ahn:2002up}
M.~H.~Ahn {\it et al.}  [K2K Collaboration],
Phys.\ Rev.\ Lett.\  {\bf 90} (2003) 041801
[arXiv:hep-ex/0212007].

\bibitem{solar}
B.~T.~Cleveland {\it et al.},
Astrophys.\ J.\  {\bf 496} (1998) 505;\\
J.~N.~Abdurashitov {\it et al.}  [SAGE Collaboration],
Phys.\ Rev.\ C {\bf 60} (1999) 055801 [arXiv:astro-ph/9907113];\\
W.~Hampel {\it et al.}  [GALLEX Collaboration],
Phys.\ Lett.\ B {\bf 447} (1999) 127;\\
S.~Fukuda {\it et al.}  [Super-Kamiokande Collaboration],
Phys.\ Rev.\ Lett.\  {\bf 86} (2001) 5651 [arXiv:hep-ex/0103032];\\
Q.~R.~Ahmad {\it et al.}  [SNO Collaboration],
Phys.\ Rev.\ Lett.\  {\bf 87} (2001) 071301
[arXiv:nucl-ex/0106015].

\bibitem{reactor}
K.~Eguchi {\it et al.}  [KamLAND Collaboration],
Phys.\ Rev.\ Lett.\  {\bf 90} (2003) 021802
[arXiv:hep-ex/0212021].

\bibitem{lsnd}
C.~Athanassopoulos {\it et al.}  [LSND Collaboration],
Phys.\ Rev.\ Lett.\  {\bf 81} (1998) 1774 [arXiv:nucl-ex/9709006];\\
A.~Aguilar {\it et al.}  [LSND Collaboration],
Phys.\ Rev.\ D {\bf 64} (2001) 112007
[arXiv:hep-ex/0104049].

\bibitem{boone}
I.~Stancu {\it et al.}  [MiniBooNE collaboration],
FERMILAB-TM-2207.

\bibitem{neutrino_osc}
B.~Pontecorvo,
Sov.\ Phys.\ JETP {\bf 6} (1957) 429 [Zh.\ Eksp.\ Teor.\ Fiz.\  {\bf
33} (1957) 549];\\
Z.~Maki, M.~Nakagawa and S.~Sakata,
Prog.\ Theor.\ Phys.\  {\bf 28} (1962) 870;\\
B.~Pontecorvo,
Sov.\ Phys.\ JETP {\bf 26} (1968) 984 [Zh.\ Eksp.\ Teor.\ Fiz.\  {\bf
53} (1967) 1717];\\
V.~N.~Gribov and B.~Pontecorvo,
Phys.\ Lett.\ B {\bf 28} (1969) 493.

\bibitem{chooz}
M.~Apollonio {\it et al.}  [CHOOZ Collaboration],
Phys.\ Lett.\ B {\bf 466} (1999) 415 [arXiv:hep-ex/9907037];\\
M.~Apollonio {\it et al.}  [CHOOZ Collaboration],
Eur.\ Phys.\ J.\ C {\bf 27} (2003) 331
[arXiv:hep-ex/0301017].

\bibitem{Apollonio:2002en}
M.~Apollonio {\it et al.},
arXiv:hep-ph/0210192.

\bibitem{betabeams_moriond} Workshop on ``Radioactive beams
for nuclear physics and neutrino physics'' 37$^{th}$ Rencontre de
Moriond, Les Arcs (France) March 17-22nd, 2003;
http://moriond.in2p3.fr/radio/index.html.


\bibitem{Burguet-Castell:2003vv}
J.~Burguet-Castell, D.~Casper, J.~J.~Gomez-Cadenas, P.~Hernandez and F.~Sanchez,
arXiv:hep-ph/0312068.

\bibitem{Terranova:2004hu}
F.~Terranova, A.~Marotta, P.~Migliozzi and M.~Spinetti,
arXiv:hep-ph/0405081.

\bibitem{Bouchez:2003fy}
J.~Bouchez, M.~Lindroos and M.~Mezzetto,
arXiv:hep-ex/0310059.

\bibitem{Gomez-Cadenas:2001eu}
J.~J.~Gomez-Cadenas {\it et al.}  [CERN working group on Super Beams
                  Collaboration],
arXiv:hep-ph/0105297.

\bibitem{allSB}
H.~Minakata and H.~Nunokawa,
Phys.\ Lett.\ B {\bf 495} (2000) 369 [arXiv:hep-ph/0004114];\\
V.~D.~Barger, S.~Geer, R.~Raja and K.~Whisnant,
Phys.\ Rev.\ D {\bf 63} (2001) 113011 [arXiv:hep-ph/0012017];\\
V.~D.~Barger {\it et al.},
arXiv:hep-ph/0103052;\\
H.~Minakata and H.~Nunokawa,
JHEP {\bf 0110} (2001) 001 [arXiv:hep-ph/0108085];\\
P.~Huber, M.~Lindner and W.~Winter,
Nucl.\ Phys.\ B {\bf 645} (2002) 3 [arXiv:hep-ph/0204352];\\
G.~Barenboim {\it et al.},
arXiv:hep-ex/0206025.

\bibitem{Burguet-Castell:2001ez}
J.~Burguet-Castell, M.~B.~Gavela, J.~J.~Gomez-Cadenas, P.~Hernandez and O.~Mena,
Nucl.\ Phys.\ B {\bf 608} (2001) 301 [arXiv:hep-ph/0103258];\\
J.~Burguet-Castell and O.~Mena,
arXiv:hep-ph/0108109.

\bibitem{Minakata:2001qm}
H.~Minakata and H.~Nunokawa,
JHEP {\bf 0110} (2001) 001 [arXiv:hep-ph/0108085].

\bibitem{Fogli:1996pv}
G.~L.~Fogli and E.~Lisi,
Phys.\ Rev.\ D {\bf 54} (1996) 3667
[arXiv:hep-ph/9604415].

\bibitem{Barger:2001yr}
V.~Barger, D.~Marfatia and K.~Whisnant,
Phys.\ Rev.\ D {\bf 65} (2002) 073023 [arXiv:hep-ph/0112119].

\bibitem{Donini:2003vz}
A.~Donini, D.~Meloni and S.~Rigolin,
arXiv:hep-ph/0312072.

\bibitem{Jung:1999jq}
C.~K.~Jung,
arXiv:hep-ex/0005046.

\bibitem{Zucchelli:sa}
P.~Zucchelli,
Phys.\ Lett.\ B {\bf 532} (2002) 166.

\bibitem{Cervera:2000kp}
A.~Cervera, A.~Donini, M.~B.~Gavela, J.~J.~Gomez Cadenas, P.~Hernandez, O.~Mena and S.~Rigolin,
Nucl.\ Phys.\ B {\bf 579} (2000) 17
[Erratum-ibid.\ B {\bf 593} (2001) 731]
[arXiv:hep-ph/0002108].

\bibitem{betadecays}
L.P. Ekstr\"{o}m and R.B. Firestone, WWW Table of Radioactive Isotopes,
from http://ie.lbl.gov/toi/index.htm

\bibitem{gilardoni}
S.~Gilardoni, to be published as CERN Thesis. \\
S.~Gilardoni, G.~Grawer, G.~Maire, J.~M.~Maugain, S.~Rangod and F.~Voelker,
J.\ Phys.\ G {\bf 29} (2003) 1801.

\bibitem{Zeller:2003ey}
G.~P.~Zeller,
arXiv:hep-ex/0312061.

\bibitem{Casper:2002sd}
D.~Casper,
Nucl.\ Phys.\ Proc.\ Suppl.\  {\bf 112} (2002) 161
[arXiv:hep-ph/0208030].

\bibitem{Serreau:2004kx}
J.~Serreau and C.~Volpe,
arXiv:hep-ph/0403293.

\bibitem{lipari}
P.~Lipari, private communication;\\ P.~Lipari, M.~Lusignoli and
F.~Sartogo,
Phys.\ Rev.\ Lett.\  {\bf 74}, 4384 (1995)
[arXiv:hep-ph/9411341].


\bibitem{Mezzetto:2003mm}
M.~Mezzetto,
J.\ Phys.\ G {\bf 29} (2003) 1781
[arXiv:hep-ex/0302005].

\bibitem{jhf}
Y.~Itow {\it et al.},
arXiv:hep-ex/0106019.

\bibitem{silver}
A.~Donini, D.~Meloni and P.~Migliozzi,
Nucl.\ Phys.\ B {\bf 646} (2002) 321
[arXiv:hep-ph/0206034];\\
D.~Autiero {\it et al.},
Eur.\ Phys.\ J.\ C {\bf 33} (2004) 243
[arXiv:hep-ph/0305185].



\end{thebibliography}
\end{document}